\documentclass[draft,oneside]{cimsart}

 \newtheorem{lemma}[theorem]{Lemma}
\newtheorem{corollary}[theorem]{Corollary}

\newcommand{\ns}[1]{\hbox{$\!\!\!\!\!#1\!\!\!\!\!$}}

{\end{veqnarray}%
}

\def\d{\displaystyle} \def\t{\textstyle} \def\scr{\scriptstyle}
\def\dx{{\it\Delta}x}  
\def\<{\langle} \def\>{\rangle}

\def\bl{{\hbox{$[$}}} \def\br{{\hbox{$]$}}}

\def\A{{\hbox{\tiny A}}} \def\B{{\hbox{\tiny B}}}

\def\out{{\hbox{\tiny out}}} \def\inn{{\hbox{\tiny in}}}

\def\dint{\hbox{$\d\int\!\!\!\!\int$}}

\def\law{\stackrel{\hbox{\tiny law}}{=}}

\newfont{\Bbb}{msbm10 at 11pt} \newfont{\bbb}{msbm10 at 8pt}

\def\R{\hbox {\Bbb R}} \def\N{\hbox {\Bbb N}}

\authorheadline{E AND VANDEN EIJNDEN} \titleheadline{STATISTICAL
  THEORY FOR BURGERS EQUATION}

\begin{document}

\title{Statistical Theory \\
  for the Stochastic Burgers Equation \\
  in the Inviscid Limit}

\author{Weinan  E\\
  \affil Courant Institute\\
  \\ and \\
  \\ Eric Vanden Eijnden\\
  \affil Courant Institute\\
  }

\maketitle

\begin{abstract}
  A statistical theory is developed for the stochastic Burgers
  equation in the inviscid limit. Master equations for the probability
  density functions of velocity, velocity difference and velocity
  gradient are derived.  No closure assumptions are made. Instead
  closure is achieved through a dimension reduction process, namely
  the unclosed terms are expressed in terms of statistical quantities
  for the singular structures of the velocity field, here the shocks.
  Master equations for the environment of the shocks are further
  expressed in terms of the statistics of singular structures on the
  shocks, namely the points of shock generation and collisions.  The
  scaling laws of the structure functions are derived through the
  analysis of the master equations.  Rigorous bounds on the decay of
  the tail probabilities for the velocity gradient are obtained using
  realizability constraints.  We also establish that the probability
  density function $Q(\xi)$ of the velocity gradient decays as
  $|\xi|^{-7/2}$ as $\xi \to -\infty$.
\end{abstract}

\newpage
\tableofcontents
\newpage

\section{Introduction}
\label{sec:1}

Consider the randomly forced Burgers equation:
\begin{veqnarray}
  \label{eq:1.1}
  u_t+u u_x =\nu u_{xx} +f,
\end{veqnarray}
where $f(x,t)$ is a zero-mean, Gaussian, statistically homogeneous,
and white-in-time random process with covariance
\begin{veqnarray}
  \nonumber \< f(x,t) f(y,s)\> = 2 B(x-y)\delta(t-s),
\end{veqnarray}
and $B(x)$ is a smooth function. In the language of stochastic
differential equation (\ref{eq:1.1}) must be interpreted as

\begin{veqnarray}
  \nonumber du =(-u u_x +\nu u_{xx}) dt+dW(x,t),
\end{veqnarray}
where $W(x,t)$ is a Wiener process with covariance
\begin{veqnarray}
  \nonumber \< W(x,t)W(y,s)\>=2B(x-y)\min(t,s).
\end{veqnarray}
We will be interested in the statistical behavior of the stationary
states (invariant measures) of (\ref{eq:1.1}) if they exist, or the
transient states with possibly random initial data.  There are two
main reasons for considering this problem. The first is that
(\ref{eq:1.1}) and its multi-dimensional version are among the
simplest nonlinear models in non-equilibrium statistical mechanics. As
such they serve as qualitative models for a wide variety of problems
including charge density waves \cite{feig83}, vortex lines in high
temperature super-conductors \cite{blfege94}, dislocations in
disordered solids and kinetic roughening of interfaces in epitaxial
growth \cite{krsp92}, formation of large-scale structures in the
universe \cite{shze89,vedufr94}, etc. The connection between these
problems and (\ref{eq:1.1}) can be understood as follows. Consider an
elastic string in a random potential $V(x,s)$.  The string is assumed
to be {\em directed} in the sense that there is a time-like direction,
assumed to be~$s$, such that the configuration of the string is a
graph over the $s$-axis. Let $Z(x,t)$ be the partition function for
the configurations of the string in the interval $0\le s\le t$, pinned
at position~$x$ at time~$t$:
\begin{veqnarray}
  \nonumber Z(x,t) = \left.\left\< {\rm e}^{-\beta \int_0^t ds\ 
        V(w(s),s)} \right|w(t)=x\right\>,
\end{veqnarray}
where $\<\cdot|w(t)=x\>$ denotes the expectation over all Brownian
paths $w(\cdot)$ such that $w(t)=x$, $\beta=1/kT$, $k$ is the
Boltzmann constant, $T$ is the temperature. The free energy,
$\varphi(x,t)=\ln Z(x,t)$ then satisfies
\begin{veqnarray}
  \nonumber \varphi_t +{\t\frac{1}{2}} |\nabla \varphi|^2 = \nu \Delta
  \varphi +V,
\end{veqnarray}
where $\nu=kT$. This is the well-known Kardar-Parisi-Zhang equation
\cite{kapazh86}. In one dimension if we let $u=\varphi_x$, $f=V_x$, we
obtain (\ref{eq:1.1}).

The second reason for studying (\ref{eq:1.1}) is in some sense a
technical one. For a long time, (\ref{eq:1.1}) has served as the
benchmark for field-theoretic techniques such as the direct
interaction approximation or the renormalization group methods,
developed for solving the problem of hydrodynamic turbulence. This
role of (\ref{eq:1.1}) is made more evident by the recent flourish of
activities introducing fairly sophisticated techniques in field theory
to hydrodynamics \cite{pol95,bomepa95,gumi96}.  In this context
(\ref{eq:1.1}) is often referred to as Burgers turbulence. Since the
phenomenology of the so-called Burgers turbulence is far simpler than
that of real turbulence, one hopes that exact results can be obtained
which can then be used to benchmark the methods. However so far our
experience has proved otherwise: The problem of Burgers turbulence is
complicated enough that a wide variety of predictions have been made
as a consequence of
the wide variety of techniques used \cite{ave95b,ave95a,avrye94,bafako97,%
bol97,bol98,bome96,chya95a,chya95b,chya96,ekhma97,eva99a,got94,gokr93,gokr98,%
gumi96,gusiau97,kra99,pol95,ryav99,sin92,sin96,vedufr94}.

The main purpose of the present paper is to clarify this situation and
to obtain exact results that are expected for Burgers turbulence. Along
the way we will also develop some technical aspects that we believe
will be useful for other problems. The main issues that interest us
are the scaling of the structure functions and the asymptotic behavior
of the probability density functions (PDF) in the inviscid limit. The
former is well-understood heuristically but we will derive the results
from self-consistent asymptotics on the master equation. The latter is
at the moment very controversial, and we hope to settle the
controversy by deriving exact results on the asymptotic behavior for
the PDFs.

From a technical point of view we insist on working with the master
equation and making no closure assumptions. The unclosed term is
expressed in terms of the statistics of singular dissipative
structures of the field, here the shocks.  We then derive a master
equation for the statistics of the environment of the shocks by
relating them to the singular structures on the shocks, namely the
points of shock creation and collisions. These are then amenable to
local analysis. In this way we achieve closure through dimension
reduction.  We then extract information on the asymptotic behavior of
PDFs using realizability constraints and self-consistent asymptotics.
We certainly hope that this philosophy will be useful for other
problems.

One main issue that will be addressed in this paper is the behavior of
the PDF of the velocity gradient. Assuming statistical homogeneity,
let $Q^\nu(\xi,t)$ be the PDF of $\xi=u_x$.  $Q^\nu$ satisfies
\begin{veqnarray}
  \nonumber Q^\nu_t=\xi Q^\nu +\bigl(\xi^2 Q^\nu\bigr)_\xi +B_1
  Q^\nu_{\xi\xi} -\nu\bigl( \< \xi_{xx}|\xi\> Q^\nu\bigr)_\xi,
\end{veqnarray}
where $B_1=-B_{xx}(0)$, $\<\xi_{xx}|\xi\>$ is the average of
$\xi_{xx}$ conditional on $\xi$. This equation is unclosed since the
explicit form of the last term, representing the effect of the
dissipation, is unknown. We are interested in $Q^\nu$ at the inviscid
limit:
\begin{veqnarray}
  \nonumber \underline{Q}(\xi,t)=\lim_{\nu\to0} Q^\nu(\xi,t).
\end{veqnarray}
In order to derive an equation for $\underline{Q}$, one needs to
evaluate
\begin{veqnarray}
  \nonumber F(\xi,t)= -\lim_{\nu\to0} \nu\bigl( \< \xi_{xx}|\xi\>
  Q^\nu\bigr)_\xi.
\end{veqnarray}
This is where the difficulty arises.

Remembering that $u_x(x,t)=\lim_{\dx\to0}(u(x+\dx,t)-u(x,t))/\dx$, the
procedure outlined above pertains to the process of first taking the
limit as $\dx\to0$, then the limit as $\nu\to0$. It is natural to
consider also the other situation when the limit $\nu\to0$ is taken
first. In this case the limiting form of (\ref{eq:1.1}), (notice that
we now write the nonlinear term in a conservative form),
\begin{veqnarray}
  \label{eq:1.2}
  u_t+{\t\frac{1}{2}}\bigl( u^2\bigr)_x = f,
\end{veqnarray}
has to be interpreted in a weak sense by requiring
\begin{veqnarray}
  \label{eq:1.3}
  \dint dxdt\ \Bigl\{ u\varphi_t +{\t\frac{1}{2}}u^2\varphi_x +
  f\varphi\Bigr\} =0,
\end{veqnarray}
for all compactly supported smooth functions $\varphi$. The solutions
$u$ satisfying (\ref{eq:1.3}) are called weak solutions. In order to
ensure a well-defined dynamics for (\ref{eq:1.2}), i.e. existence and
uniqueness of solutions with given initial data, an additional entropy
condition has to be imposed on weak solutions. This amounts to
requiring
\begin{veqnarray}
  \label{eq:1.4}
  u(x{\scr+},t)\le u(x{\scr-},t),
\end{veqnarray}
for all $(x,t)$. The entropy condition (\ref{eq:1.4}) is the effect of
the viscous term in the inviscid limit. There is a huge mathematical
literature on (\ref{eq:1.2}) for the deterministic case. Standard
references are \cite{lax57,lax73,smo83}. The random case was studied
recently in the paper by E, Khanin, Mazel and Sinai \cite{ekhma99}.

The first step in the present paper is to derive master equations for
single and multi-point statistics of $u$ satisfying (\ref{eq:1.2}). In
particular, we derive an equation for the PDF of
$\eta(x,y,t)=(u(x+y)-u(x,t))/y$, $Q^\delta(\eta,x,t)$.  We are
interested in
\begin{veqnarray}
  \nonumber Q(\xi,t)=\lim_{x\to0}Q^\delta(\xi,x,t).
\end{veqnarray}
One natural question is whether
\begin{veqnarray}
  \label{eq:1.4.1}
  Q=\underline Q.
\end{veqnarray}
We will present very strong argument that (\ref{eq:1.4.1}) holds for
generic initial data and for the type of forces described after
(\ref{eq:1.1}).  Some of our results also apply to $Q$ for the more
general case when it is possibly different from $\underline Q$.

The issue now reduces to the evaluation or approximation of
$F(\xi,t)$. Several different proposals have been made, each leads at
statistical steady state ($Q_t=0$) to an asymptotic expression of the
form
\begin{veqnarray}
  \nonumber Q(\xi) \sim \left\{
  \begin{array}{ll}
    {\d C_-|\xi|^{-\alpha}}& {\rm as} \ \ 
    \xi\to-\infty,\\[6pt]
    {\d C_+\xi^\beta {\rm e}^{-\xi^3/(3B_1)}\qquad}& {\rm as} \ \ 
    \xi\to+\infty,
  \end{array}\right.
\end{veqnarray}
but with a variety of values for the exponents $\alpha$ and $\beta$
(here the $C_\pm$'s are constants). By invoking the operator product
expansion, Polyakov \cite{pol95} suggested that $F=aQ+b\xi Q$, with
$a=0$ and $b=-1/2$.  This leads to $\alpha=5/2$ and $\beta=1/2$.
Boldyrev \cite{bol97,bol98} considered the same closure with $-1\leq
b\leq 0$, which gives $2\leq \alpha\leq 3$ and $\beta=1+b$.  Based on
heuristic arguments, Bouchaud and M{\'e}zard \cite{bome96} introduced
a Langevin equation for the local slope of the velocity, which gives
$2\le \alpha\le 3$, $\beta=0$. The instanton analysis
\cite{bafako97,fakole96,gumi96} predicts the right tail of $Q$ without
giving a precise value for $\beta$, and it does not give any specific
prediction for the left tail.  E~{\it et~al.}  \cite{ekhma97} made a
geometrical evaluation of the effect of $F$, based on the observation
that large negative gradients are generated near shock creation.
Their analysis gives a rigorous upper-bound for $\alpha$: $\alpha \le
7/2$.  In~\cite{ekhma97}, it was claimed that this bound is actually
reached, i.e., $\alpha=7/2$.  Finally Gotoh and Kraichnan
\cite{gokr98} argued that the viscous term is negligible to leading
order for large ${|\xi|}$, i.e.  $F\approx 0$ for $|\xi|\gg
B_1^{1/3}$, giving rise to $\alpha=3$ and $\beta=1$.

In this paper, we will first give a more detailed proof of the bound
\begin{veqnarray}
  \nonumber \alpha>3,
\end{veqnarray}
announced in \cite{eva99a}. We will then show that
\begin{veqnarray}
  \nonumber \alpha=7/2,
\end{veqnarray}
by studying the master equation for the environment of the shocks.

An important reason behind the success of this program is that local
behavior near the most singular structures (here the points of shock
creation) is understood.  Points of shock creation were isolated in
\cite{ekhma99} as a mechanism for obtaining large negative values of
$u_x$ and resulted in the prediction that $\alpha=7/2$. From this
point of view, the present paper provides the missing step
establishing the fact that points of shock creation provide the
leading order contribution to the left tail of $Q$.

For the most part, our working assumption will be that solutions of
(\ref{eq:1.2}) are piecewise smooth. Shock path are smooth except at
the points of collision. In particular, shocks are created at zero
amplitude and the shock strength adds up at collision. For transient
states these statements follow from standard results for the
deterministic case under appropriate conditions for the initial data
\cite{liu99}. For stationary states, these results are of the type
established in \cite{ekhma99} under a non-degeneracy condition for
the forcing.

Before ending this introduction, we make some remarks about notations
and nomenclature.  In analogy with fluid mechanics $u$ will be
referred to as the velocity field.  We will denote the multi-point
PDFs of $u(\cdot,t)$ as $Z^\nu(u_1,x_1,\ldots,u_n,x_n,t)$, i.e.
\begin{veqnarray}
  \nonumber
  &&\hbox{Prob} (a_1\le u(x_1,t)\le b_1,\ldots,a_n\le u(x_n,t)\le b_n)\\
  &=& \int_{a_1}^{b_1}\!\!du_1\cdots \int_{a_n}^{b_n}\!\!du_n\ 
  Z^\nu(u_1,x_1,\ldots,u_n,x_n,t).
\end{veqnarray}
The superscript $\nu$ refers to the viscous case, $\nu>0$. In the
inviscid limit we will denote the multi-point PDFs of $u(\cdot,t)$ by
$Z(u_1,x_1\ldots,u_n,x_n,t)$. Statistical stationary values will be
denoted by the subscript $\infty$, e.g. $Z^\nu_\infty$ or $Z_\infty$.
We reserve the special notations $R^\nu(u,x,t)=Z^\nu(u,x,t)$ and
$R(u,x,t)=Z(u,x,t)$ for the one-point PDF of $u(x,t)$.  $Q^\nu(\xi,t)$
denotes the PDF of $\xi(x,t)$, and its inviscid limit is $Q(\xi,t)$.
Statistical symmetries, or equality in law, will be denoted by $\law$.
Two of such symmetries will be repeatedly used. The first is
statistical homogeneity,
\begin{veqnarray}
  \label{eq:1.5}
  u(x,t)\law u(x+y,t) \quad \hbox{for all } y.
\end{veqnarray}
(\ref{eq:1.5}) holds if for all measurable, real-valued functional
$\Phi(u(\cdot,t))$ we have
\begin{veqnarray}
  \nonumber \< \Phi(u(\cdot,t))\> = \< \Phi(u(\cdot+y,t))\> \quad
  \hbox{for all } y.
\end{veqnarray}
Note that from (\ref{eq:1.1}) and the assumptions on~$f$, it follows
that (\ref{eq:1.5}) holds for all times if $u_0(x)\law u_0(x+y)$,
where $u_0(\cdot)=u(\cdot,0)$ is the initial velocity field. Also, if
a statistical steady state exists and is unique, then it satisfies
(\ref{eq:1.5}). The second symmetry will be referred to as statistical
parity invariance, and is related to the invariance of (\ref{eq:1.1})
under the transformation $x\to-x$, $u\to-u$. This implies that
\begin{veqnarray}
  \label{eq:1.6}
  u(x,t)\law -u(-x,t),
\end{veqnarray}
or
\begin{veqnarray}
  \nonumber \< \Phi(u(\cdot,t))\> = \< \Phi(-u(-(\cdot),t))\>.
\end{veqnarray}
(\ref{eq:1.6}) holds for all times if $u_0(x)\law -u_0(-x)$.
(\ref{eq:1.6}) is also satisfied at statistical steady state.

Even though some of our results are formulated in terms of lemmas and
theorems, the emphasize here is on the ability to calculate things
rather than rigor.  
We will assume that the initial distribution
$Z_0(u_1,x_1,\ldots,u_n,x_n)= Z(u_1,x_1,\ldots,u_n,x_n,0)$ is
concentrated on $D(I)$, the space of functions defined on $I\subset
\R$ admitting only jumps discontinuities:
\begin{veqnarray}
  \nonumber D(I)=\{u(x,t), u(x{\scr\pm},t) \,\mbox{exists for all}\,
  x, \hbox{ with } u(x{\scr+},t)\le u(x{\scr-},t)\}.
\end{veqnarray}
Note that the existence of stationary states is only established when
$I$ is a finite interval on $\R$ \cite{ekhma99}.  We will also assume
that the initial velocity field $u_0(\cdot)=u(\cdot,0)$ is independent
of $\nu$ and statistically independent of $f(\cdot)$.

Finally, unspecified integration ranges are meant to be
$(-\infty,+\infty)$.

\section{Velocity and velocity differences}
\label{sec:3}

Let $u^\nu(x,t,f,u_0)$ be the solution of (\ref{eq:1.1}) with forcing
$f$, initial data $u_0$. It follows from standard results (for the
deterministic case, see \cite{lax57,lax73}) that for fixed~$t$
\begin{veqnarray}
  \nonumber u^\nu(\cdot,t,f,u_0)\to u(\cdot,t,f,u_0) \quad \hbox{as }
  \nu\to0,
\end{veqnarray}
$P\times P_0$-almost surely, in $L^2(I)$, where $I$ is the interval on
which (\ref{eq:1.1}) is studied. Hence
\begin{veqnarray}
  \nonumber Z^\nu \to Z,
\end{veqnarray}
weakly. For the statistically stationary states, it was established in
\cite{ekhma99} that, weakly, we have
\begin{veqnarray}
  \nonumber Z_\infty^\nu \to Z_\infty.
\end{veqnarray}
Therefore to study the inviscid limit of statistical quantities of
$u$, such as the PDFs of $u$ and $\delta u(x,y,t)=u(x+y,t)-u(y,t)$, it
is enough to consider (\ref{eq:1.2}).

It is useful, however, to write this equation in a modified form more
convenient for calculations.  To this end, let
\begin{veqnarray}
  \nonumber u_\pm(x,t)=u(x\pm,t).
\end{veqnarray}
For any function $g(u)=g(u(x,t))$, we define two average functions
\begin{veqnarray}
  \label{eq:3.1}
  \bl g(u)\br_\A &=& {\t \frac{1}{2}} (g(u_+)+g(u_-))\\
  \bl g(u)\br_\B &=& \int_0^1\!\! d\beta\ g(u_-+\beta(u_+-u_-)).
\end{veqnarray}
It is shown in \cite{vol67} that (\ref{eq:1.2}) is equivalent to
\begin{veqnarray}
  \label{eq:3.3}
  u_t+\bl u\br_\A u_x = f,
\end{veqnarray}
or, in the language of stochastic differential equations,
\begin{veqnarray}
  \label{eq:3.3b}
  du=-\bl u\br_\A u_xdt +dW(x,t).
\end{veqnarray}
(\ref{eq:3.1}) assigns an unambiguous meaning to the quantity $\bl
u\br_\A u_x$ when the solutions of (\ref{eq:3.3}) develop shocks.
Moreover the following chain and product rules hold
\begin{veqnarray}
  \label{eq:3.4}
  g_x(u)=\bl g_u(u)\br_\B u_x,
\end{veqnarray}
\begin{veqnarray}
  \label{eq:3.5}
  (g(u)h(u))_x=\bl g(u)\br_\A h_x(u) + \bl h(u)\br_\A g_x(u).
\end{veqnarray}
Similar rules apply for derivatives in $t$. 
As an example of these rules, we have that
the integral of the term $\bl u\br_\A u_x $ across a shock located at
$x=y$ is given by (using $\bl u\br_\A=\bl u\br_\B$)
\begin{veqnarray}
  \nonumber \int_{y-}^{y+}\!\!dx \ \bl u\br_\A u_x =
  \int_{y-}^{y+}\!\!dx \  {\t \frac{1}{2}}(u^2)_x = {\t \frac{1}{2}}
  (u_+^2-u_-^2).
\end{veqnarray}

It is also convenient to define
\begin{veqnarray}
  \nonumber
  \bar u(x,t)&=&{\t \frac{1}{2}} (u_+(x,t)+u_-(x,t)), \\
  \nonumber s(x,t)&=&u_+(x,t)-u_-(x,t).
\end{veqnarray}
In terms of $(\bar u,s)$, the two averages in (\ref{eq:3.1}) are
\begin{veqnarray}
  \nonumber \bl g(u)\br_\A &=&
  {\t \frac{1}{2}} (g(\bar u+s/2)+g(\bar u-s/2))\\
  \nonumber \bl g(u)\br_\B &=& \int_{-1/2}^{1/2} \!\!d\beta\ g(\bar u
  +\beta s).
\end{veqnarray}
We will denote by $\{y_j\}$ the set of shock positions at time $t$.
Hence the set
\begin{veqnarray}
  \nonumber \{(y_j,\bar u(y_j,t),s(y_j,t))\},
\end{veqnarray}
quantifies the shocks at time~$t$.

\subsection{Master equation for the inviscid Burgers equation}
\label{sec:3.1}

We now turn to $R(u,x,t)$, the one-point PDF of the solution of
(\ref{eq:3.3}).  We have

\begin{theorem}
\label{th:3.1}
$R$ satisfies
\begin{veqnarray}
  \label{eq:3.6}
  R_t=-u R_x -\int du'\ K(u-u') R_x(u',x,t)+ B_0 R_{uu}+G,
\end{veqnarray}
where $B_0=B(0)$, $K(u)=(H(u)-H(-u))/2$, $H(\cdot)$ is the Heaviside
function and $G(u,x,t)$ is given by
\begin{veqnarray}
  \label{eq:3.7}
  G(u,x,t)=\left( \int_{-\infty}^0\!\! d s \int_{u+s/2}^{u-s/2}\!\!
    d\bar u \ (u-\bar u) \varrho(\bar u,s,x,t)\right)_u.
\end{veqnarray}
Here $\varrho(\bar u,s,x,t)$ is defined such that $\varrho(\bar
u,s,x,t)d \bar u ds dx$ gives the average number of shocks in $[x,x+dx)$
with $ \bar u(y,t) \in [\bar u, \bar u+d\bar u)$ and $s(y,t) \in [s,
s+ds)$, where $y\in[x,x+dx)$ is the shock location.

\end{theorem}

\demo{Remark 1} $G$ can be referred to as a {\em dissipative
  anomaly}. It can be written more explicitly as
\begin{veqnarray}
  \label{eq:3.8}
  G(u,x,t)&=&\frac{1}{2} \int_{-\infty}^0\!\!  d s
  \ s \bigl(\varrho(u-s/2,s,x,t)+\varrho(u+s/2,s,x,t)\bigr)\\
  &-&\int_{-1/2}^{1/2}\!\!d\beta\int_{-\infty}^0\!\!  ds \ s
  \varrho(u+\beta s,s,x,t).
\end{veqnarray}
\enddemo \demo{Remark 2} $\varrho (\bar u,s,x,t)$ can be
equivalently defined as
\begin{veqnarray}
  \nonumber \varrho (\bar u,s,x,t)= \int \frac{d\lambda
    d\mu}{(2\pi)^2}\ {\rm e}^{i\lambda \bar u+i\mu s} \,\hat
  \varrho(\lambda,\mu,x,t).
\end{veqnarray}
where
\begin{veqnarray}
  \nonumber \hat \varrho (\lambda,\mu,x,t)= \biggl\< \sum_j{\rm
    e}^{-i\lambda \bar u(y_j,t)-i\mu s(y_j,t)} \delta(x-y_j)\biggr\>.
\end{veqnarray}
\enddemo

~

(\ref{eq:3.6}), (\ref{eq:3.7}) do not require statistical homogeneity.
For homogeneous situations, these equations simplify. Since $R_x=0$
(\ref{eq:3.6}) reduces to
\begin{veqnarray}
  \label{eq:3.6.b}
  R_t=B_0 R_{uu}+G,
\end{veqnarray}
where $G(u,x,t)=G(u,t)$.  Since the shock characteristics are
independent of its location, we have

\begin{veqnarray}
  \nonumber \varrho(\bar u,s,x,t)=\rho S(\bar u,s,t)
\end{veqnarray}
where $\rho=\rho(t) $ is the number density of shocks, $S(\bar u,s,t)$
is the PDF of $(\bar u(y_0,t),s(y_0,t))$, conditional on the property
that $y_0$ is a shock position (because of the statistical
homogeneity, $y_0$ is a dummy variable).  Thus, (\ref{eq:3.7}) reduces
to
\begin{veqnarray}
  \label{eq:3.10}
  G(u,t)=\rho\left( \int_{-\infty}^0\!\!  d s \int_{u+s/2}^{u-s/2}\!\!
    d\bar u \ (u-\bar u) S(\bar u,s,t)\right)_u.
\end{veqnarray}

At statistical stationary state, $R_t=0$ in (\ref{eq:3.6.b}), and one
readily verifies from this equation that
\begin{veqnarray}
  \label{3.12.b}
  R_\infty(u)=\frac{\rho}{2B_0} \int_{-\infty}^0\!\!  d s
  \int_{u+s/2}^{u-s/2}\!\! d\bar u \ \left(\frac{s^2}{4}-(u-\bar
    u)^2\right) S_\infty(\bar u,s),
\end{veqnarray}
Clearly, $R_\infty\ge 0$ since $S_\infty(\bar u,s)\ge0$. In addition,
by direct computation we obtain
\begin{veqnarray}
  \nonumber \int du \ R_\infty(u)=-\frac{\rho}{12B_0}\< s^3\>.
\end{veqnarray}
Thus, from the requirement that $R_\infty$ be normalized to unity, we
get
\begin{veqnarray}
  \label{eq:3.15b}
  B_0=-\frac{\rho}{12} \< s^3\>.
\end{veqnarray}

\begin{proof}[Proof of Theorem \ref{th:3.1}]
  Let $\theta(\lambda,x,t) = {\rm e}^{-i\lambda u(x,t)}$.  $\<
  \theta\>$ is the characteristic function of $u(x,t)$ and $R$ is
  given by
\begin{veqnarray}
  \nonumber R(u,x,t)=\int \frac{d\lambda}{2\pi} {\rm e}^{i\lambda u}
  \<\theta(\lambda,x,t)\>.
\end{veqnarray}
Using Ito calculus,
\begin{veqnarray}
  \nonumber dW(x,t)dW(y,t)=2B(x-y)dt,
\end{veqnarray}
it follows from (\ref{eq:3.3b}), (\ref{eq:3.4}), (\ref{eq:3.5}) that
\begin{veqnarray}
  \nonumber d\theta = (-i\lambda\bl u\br_\A u_x\bl\theta\br_\B
  -\lambda^2 B_0\theta)dt -i\lambda \theta dW(x,t).
\end{veqnarray}
Thus
\begin{veqnarray}
  \nonumber \<\theta\>_t = i\lambda \<\bl u\br_\A u_x\bl\theta\br_\B\>
  -\lambda^2 B_0 \<\theta\>.
\end{veqnarray}
Note that the terms involving the force contain $\theta$, not
$\bl\theta\br_\B$. This is because $\theta=\bl\theta\br_\B$ except
when there is a shock, and this is a set of zero probability. Of
course this argument does not apply for the convection term since
$u_x$ is infinite at shocks. To average the convective term $i\lambda
\bl u\br_\A u_x\bl\theta\br_\B$ we use
\begin{veqnarray}
  \nonumber \theta_x= -i\lambda u_x \bl\theta\br_\B
\end{veqnarray}
to get
\begin{veqnarray}
  \nonumber i\lambda \bigl\<\bl u\br_\A u_x \bl\theta\br_\B\bigr\> = -
  \bigl\<\bl u\br_\A \theta_x\bigr\> = - \bigl\< u \theta\bigr\>_x +
  \bigl\< u_x \bl\theta\br_\A\bigr\>.
\end{veqnarray}
For convenience we write this equation as
\begin{veqnarray}
  \nonumber i\lambda \bigl\<\bl u\br_\A u_x \bl\theta\br_\B\bigr\>
  =-i\<\theta\>_{x\lambda} +i\lambda^{-1}\<\theta\>_x +\bigl\< u_x
  \bigl(\bl\theta\br_\A-\bl\theta\br_\B \bigr)\bigr\>.
\end{veqnarray}
Combining these expressions gives:
\begin{veqnarray}
  \label{eq:3.11.0}
  \<\theta\>_t =-i\<\theta\>_{x\lambda}
  +i\lambda^{-1}\<\theta\>_x-\lambda^2 B_0 \<\theta\> +\hat G,
\end{veqnarray}
where $\hat G(\lambda,x,t)$ is given by
\begin{veqnarray}
  \nonumber \hat G(\lambda,x,t)=\bigl\< u_x
  \bigl(\bl\theta\br_\A-\bl\theta\br_\B \bigr) \bigr\>.
\end{veqnarray}

To proceed with the evaluation of $\hat G$, note that the only
contributions to this term are from the shocks, since
$\bl\theta\br_\A=\bl\theta\br_\B$ except at shocks. Let $\{y_j\}$
denote the positions of the shocks at time $t$.  Since
$u_x(y_j,t)=s(y_j,t)\delta(x-y_j)$ at the shocks, $\hat G$ can be
understood as
\begin{veqnarray}
  \label{eq:3.11}
  &&\hat G(\lambda,x,t)\\
  &=&\biggl\< \sum_j s_j \delta(x-y_j)
  \Bigl(\bl\theta(\lambda,y_j,t)\br_\A -\bl\theta(\lambda,y_j,t)\br_\B
  \Bigr) \biggr\>,
\end{veqnarray}
where $s_j=s(y_j,t)$.  Using (\ref{eq:3.1}),
\begin{veqnarray}
  \nonumber
  &&\hat G(\lambda,x,t)\\
  &=&\biggl\< \sum_j s_j \delta(x-y_j) {\rm e}^{-i\lambda \bar u_j}
  \biggl({\rm e}^{i\lambda s_j/2}+{\rm e}^{-i\lambda s_j/2}
  -2\int_{-1/2}^{1/2}\!\! d\beta\ {\rm e}^{-i\lambda\beta
    s_j}\biggr)\biggr\>,
\end{veqnarray}
where $\bar u_j=\bar u(y_j,t)$. Using $\varrho(\bar u,s,x,t)$ this
average is
\begin{veqnarray}
  \nonumber
  &&\hat G(\lambda,x,t)\\
  &=&\int d\bar u \int_{-\infty}^{0}\!\!ds \ \frac{s}{2}{\rm
    e}^{-i\lambda \bar u} \biggl({\rm e}^{i\lambda s/2}+{\rm
    e}^{-i\lambda s/2} -2\int_{-1/2}^{1/2}\!\! d\beta\ {\rm
    e}^{-i\lambda \beta s}\biggr) \varrho(\bar u,s,x,t),
\end{veqnarray}
Going back to the variable $u$, we get (\ref{eq:3.8}), hence
(\ref{eq:3.6}).

\end{proof}

\demo{Remark} Under the assumption of ergodicity with respect to
spatial translations, an alternative derivation of (\ref{eq:3.10}) is
to go back to (\ref{eq:3.11}) and use the equivalence between ensemble
average and spatial average.  Then
\begin{veqnarray}
  \nonumber
  &&\hat G(\lambda,t)\\
  &=&\lim_{L\to\infty}\frac{1}{2L}\int_{-L}^L\!\!dx\ \sum_j s_j
  \delta(x-y_j) \Bigl(\bl\theta(\lambda,y_j,t)\br_\A
  -\bl\theta(\lambda,y_j,t)\br_\B \Bigr)
  \biggr\>\\
  &=&\lim_{L\to\infty}\frac{N}{2L}\frac{1}{N} \sum_{j=1}^N s_j
  \Bigl(\bl\theta(\lambda,y_j,t)\br_\A- \bl\theta(\lambda,y_j,t)\br_\B
  \Bigr),
\end{veqnarray}
where $N$ is the number of shocks in the interval $[-L,L]$.  Using the
ergodicity again it follows that the sum is equal to
\begin{veqnarray}
  \nonumber
  &&\hat G(\lambda,t)\\
  &=&\rho \int d\bar u \int_{-\infty}^{0}\!\!ds \frac{s}{2}{\rm
    e}^{-i\lambda \bar u} \biggl({\rm e}^{i\lambda s/2}+{\rm
    e}^{-i\lambda s/2} -2\int_{-1/2}^{1/2}\!\! d\beta\ {\rm
    e}^{-i\lambda\beta s}\biggr) \,S(\bar u,s,t),
\end{veqnarray}
where we used $\lim_{L\to\infty} N/(2L)=\rho $. Going back to the
variable $u$, we get (\ref{eq:3.10}).  
\enddemo

\subsection{An alternative derivation}
\label{sec:3.2}

The derivation of (\ref{eq:3.6}) (or (\ref{eq:3.6.b}) for
statistically homogeneous situations) given above is rigorous, but
rather unintuitive.  In fact, the effect of the viscous term is buried
in the definition of the two averages in (\ref{eq:3.1}), and it is not
clear at all at this stage whether (\ref{eq:3.6}) arises in the limit of
the equation for $R^\nu$ as $\nu\to0$. Assuming statistical
homogeneity, recall that $R^\nu$ satisfies (see (\ref{eq:A.5}) in the
Appendix)
\begin{veqnarray}
  \nonumber R^\nu_t=B_0 R^\nu_{uu} -\nu \bigl(\< u_{xx}|u\>
  R^\nu\bigr)_{u}.
\end{veqnarray}
Here we give another, less rigorous but more intuitive, derivation of
(\ref{eq:3.6.b}) and (\ref{eq:3.10}) by working with the equation for
$R^\nu$, and calculating directly the limit of the viscous term as
$\nu\to0$. In particular, we will compute explicitly that the
dissipative anomaly (\ref{eq:3.10}) is given by
\begin{veqnarray}
  \label{eq:3.9}
  G(u,t) = -\lim_{\nu\to0} \nu \bigl(\< u_{xx} |u\> R\bigr)_u.
\end{veqnarray}
This will give strong support to the claim that $\underline
R=\lim_{\nu\to0}R^\nu$ satisfies (\ref{eq:3.6.b}).

Assuming spatial ergodicity, the average of the dissipative term can
be expressed as
\begin{veqnarray}
  \label{eq:3.12}
  \nu \< u_{xx}|u\> R^\nu
  &=&\nu \< u_{xx}(x,t) \delta(u-u(x,t))\>\\
  &=&\nu\lim_{L\to\infty}\frac{1}{2L} \int_{-L}^{L} dx\ u_{xx}(x,t)
  \delta(u-u(x,t)).
\end{veqnarray}
Clearly, in the limit as $\nu\to0$ only small intervals around the
shocks will contribute to the integral. In these intervals, boundary
layer analysis can be used to obtain an accurate approximation of
$u(x,t)$.

The basic idea is to split $u$ into the sum of an inner solution near
the shock and an outer solution away from the shock, and using
systematic matched asymptotics to construct uniform approximation of
$u$ (for details see, e.g., \cite{goxi92}).  For the outer solution,
we look for an approximation in the form of a series in $\nu$:
\begin{veqnarray}
  \nonumber u=u^\out= u_0+\nu u_1+O(\nu^2).
\end{veqnarray}
Then $u_0$ satisfies
\begin{veqnarray}
  \nonumber {u_0}_t+u_0 {u_0}_x=f,
\end{veqnarray}
i.e. Burgers equation without the dissipation term. In order to deal
with the inner solution around the shock, let $y=y(t)$ be the position
of a shock, and define the stretched variable $z=(x-y)/\nu$ and let
\begin{veqnarray}
  \nonumber u^\inn(x,t)=v\left(\frac{x-y}{\nu}+\delta,t\right),
\end{veqnarray}
where $\delta$ is a perturbation of the shock position to be
determined later.  Then, $v$ satisfies
\begin{veqnarray}
  \label{eq:3.13}
  \nu v_t +(v-\bar u+\nu \gamma)v_z =v_{zz}+\nu f,
\end{veqnarray}
where $\bar u =dy/dt$, $\gamma = d\delta/dt$ and, to $O(\nu^2)$, $\nu
f$ can be evaluated at $x=y$ (and can thus be considered as a function
of $t$ only).

We study (\ref{eq:3.13}) by regular perturbation analysis. We
look for a solution in the form
\begin{veqnarray}
  \nonumber v= v_0 +\nu v_1+O(\nu^2).
\end{veqnarray}
To leading order, from (\ref{eq:3.13}) we get for $v_0$ the equation
\begin{veqnarray}
  \nonumber (v_0-\bar u){v_0}_z ={v_0}_{zz}.
\end{veqnarray}
The boundary condition for this equation arises from the matching
condition with $u^\out=u_0+\nu u_1 +O(\nu^2)$:
\begin{veqnarray}
  \nonumber \lim_{z\rightarrow \pm\infty} v_0 = \lim_{x\to y} u_0
  \equiv \bar u\pm\frac{s}{2},
\end{veqnarray}
where $s=s(t)$ is the shock strength.  It is understood that for small
$\nu$ matching takes place for small values of $|x-y|$ and large
values of $|z|=|x-y|/\nu$. This gives
\begin{veqnarray}
  \nonumber v_0= \bar u-\frac{s}{2} \tanh \left(\frac{s z}{4}\right).
\end{veqnarray}

These results show that, to $O(\nu)$, (\ref{eq:3.12}) can be estimated
as
\begin{veqnarray}
  \nonumber \nu \< u_{xx}|u\> R^\nu =\nu\lim_{L\to\infty}\frac{N}{2L}
  \frac{1}{N} \sum_{j} \int_{\Omega_j}\!\!  dx\ u^\inn_{xx}\,
  \delta(u-u^\inn(x,t)),
\end{veqnarray}
where $\Omega_j$ is a layer centered at $y_j$ with width $\gg O(\nu)$.
Going to the stretched variable $z=(x-y)/\nu$, and taking the limit as
$L\to\infty$, we get
\begin{veqnarray}
  \nonumber \nu \< u_{xx}|u\> R^\nu =\rho \int d\bar u
  \int_{-\infty}^{0}\!\!ds \ S(\bar u,s;t)\int_{-\infty}^{+\infty}\!\!
  dz \ {v_0}_{zz}(z,t) \delta(u-v_0(z,t)),
\end{veqnarray}
where $S(\bar u,s;t)$ is the PDF of $(\bar u(y_0,t),s(y_0,t))$
conditional on $y_0$ being a shock location.  The $z$-integral can be
evaluated exactly using
\begin{veqnarray}
  \nonumber dz {v_0}_{zz}=dv_0\frac{{v_0}_{zz}}{{v_0}_z}= dv_0
  (v_0-\bar u),
\end{veqnarray}
where we used the equation $(v_0-\bar u){v_0}_z ={v_0}_{zz}$.  This
leads to
\begin{veqnarray}
  \nonumber \nu \< u_{xx} |u\> R^\nu =-\rho \int d\bar u
  \int_{-\infty}^{0}\!\!ds \ S(\bar u,s;t)\int_{\bar u+s/2}^{\bar
    u-s/2} \!\! dv_0\ (v_0-\bar u) \delta(u-v_0).
\end{veqnarray}
Hence,
\begin{veqnarray}
  \nonumber \lim_{\nu\to0}\nu \< u_{xx}|u\> R^\nu=
  -\rho\int_{-\infty}^0\!\! d s \int_{u+s/2}^{u-s/2}\!\! d\bar u \ 
  (u-\bar u) S(\bar u,s;t).
\end{veqnarray}
Using this expression in (\ref{eq:3.9}), we get (\ref{eq:3.7}).

\subsection{Computing  the anomalies} 
\label{sec:3.3}

Here we derive equations for the moments of $R$. For simplicity,
consider a statistically homogeneous case, and assume that $u(x,t)\law
-u(-x,t)$. Then $S(\bar u,s,t)=S(-\bar u,s,t)$, $G(u,t)=G(-u,t)$,
$R(u,t)=R(-u,t)$. This means that all moments of odd orders of $R$ are
zero. For the moments of even orders, we get from (\ref{eq:3.6.b}) the
following equations ($n\in\N_0$):
\begin{veqnarray}
  \nonumber \frac{d}{dt}\< u^{2n}\> = 2n(2n-1)B_0\< u^{2n-2}\>+h_{2n},
\end{veqnarray}
where $h_{2n}$ is the anomaly term
\begin{veqnarray}
  \nonumber
  h_{2n} &=& \int du \ u^{2n} G \\
  &=&-\frac{\rho}{2^{2n}(2n+1)} \Bigl(\< (2\bar u-s)^{2n}(\bar u
  +ns)\>- \< (2\bar u+s)^{2n}(\bar u -ns)\>\Bigr).
\end{veqnarray}
($h_{2n+1}=0$ by parity).  An alternative definition of $h_{2n}$ is
\begin{veqnarray}
  \nonumber h_{2n}= 2n \lim_{\nu\to0}\nu \< u^{2n-1} u_{xx}\> = -
  2n(2n-1) \lim_{\nu\to0}\nu \< u^{2n-2} u^2_x\>.
\end{veqnarray}
This gives for instance
\begin{veqnarray}
  \nonumber h_2=-2\lim_{\nu\to0}\nu \< u_x^2\> =\frac{\rho}{6} \<
  s^3\>.
\end{veqnarray}
At statistical steady state, it gives
\begin{veqnarray}
  \nonumber
  \< u^{2n}\>=\frac{C_{2n}\rho}{B_0} \Bigl(\< (2\bar u-s)^{2n+2}(\bar u
  +(n+1)s)\>- \< (2\bar u+s)^{2n+2}(\bar u -(n+1)s)\>\Bigr),
\end{veqnarray}
where $C_n=n!/2^{n+2}(n+3)!$. These expressions can also be obtained
from (\ref{3.12.b}). In particular, for $n=2$, we obtain again
(\ref{eq:3.15b}).

\subsection{Multi-point PDF}
\label{sec:3.5}

We now turn to $Z(u_1,x_1,\ldots,u_n,x_n,t)$, the multi-point PDF of
$u$. We have

\begin{theorem}
\label{th:4}
$Z$ satisfies
\begin{veqnarray}
  \label{eq:3.150}
  Z_t&=&-\sum_{p=1}^n u_pZ_{x_p}+\sum_{p,q=1}^n B(x_p-x_q)Z_{u_pu_q}\\
  &&-\sum_{p=1}^n\int du'\ K(u_p-u')
  Z_{x_p}(u_1,x_1,\ldots,u',x_p,\ldots,u_n,x_n,t)\\
  &&+\sum_{p=1}^n
  G(u_p,x_p,u_2,x_2\ldots,u_{p-1},x_{p-1},u_{p+1},x_{p+1},
  \ldots,u_n,x_n,t),
\end{veqnarray}
where $K(u)=(H(u)-H(-u))/2$, $H(\cdot)$ is the Heaviside function and
$G$ is given by
\begin{veqnarray}
  \label{eq:3.16}
  &&G(u_1,x_1,\ldots,u_n,x_n,t)\\
  &&=\biggl( \int_{-\infty}^0\!\! d s
  \int_{u_1+s/2}^{u_1-s/2}\!\! d\bar u \ (u_1-\bar u)\\[-6pt]
  &&\qquad\qquad\qquad\times \varrho(\bar
  u,s,x_1,u_2,x_2,\ldots,u_n,x_n,t)\biggr)_{u_1}.
\end{veqnarray}
Here $\varrho(\bar u,s,x_1,u_2,x_2\ldots,u_n,x_n,t)$ is defined such
that
\begin{veqnarray}
  \nonumber \varrho(\bar u,s,x_1,u_2,x_2,\ldots,u_n,x_nt)d \bar u
  dsdu_2\cdots du_n dx,
\end{veqnarray}
gives the average number of shocks in $[x_1,x_1+dx)$ with $ \bar
u(y,t) \in [\bar u, \bar u+d\bar u)$, $s(y,t) \in [s, s+ds)$ where
$y\in[x_1,x_1+dx)$ is a shock location, and with $u(x_2,t)\in [u_2,
u_2+du_2)$,\ldots, $u(x_n,t)\in [u_n, u_n+du_n)$.

\end{theorem}

We will omit the proof of Theorem~\ref{th:4} since it is a
straightforward generalization of the one given for
Theorem~\ref{th:3.1}.

~

For the two-point PDF, $Z(u_1,x_1,u_2,x_2,t)$, (\ref{eq:3.150})
becomes
\begin{veqnarray}
  \label{eq:3.15}
  Z_t&=&-u_1Z_{x_1}
  -\int du'\ K(u_1-u') Z_{x_1}(u',x_1,u_2,x_2,t)\\
  && -u_2Z_{x_2}
  - \int du'\ K(u_2-u') Z_{x_2}(u_1,x_1,u',x_2,t)\\
  &&+B_0 Z_{u_1u_1}+B_0 Z_{u_2u_2}+2B(x_1-x_2)Z_{u_1u_2}\\
  &&+G(u_1,x_1,u_2,x_2,t)+G(u_2,x_2,u_1,x_1,t).
\end{veqnarray}
Assuming statistical homogeneity, (\ref{eq:3.15}) can be simplified.
First,
\begin{veqnarray}
  \nonumber Z(u_1,y,u_2,x+y,t)=Z(u_1,0,u_2,x,t)\equiv Z(u_1,u_2,x,t).
\end{veqnarray}
Secondly,
\begin{veqnarray}
  \nonumber \varrho(\bar u,s,y,u,x+y,t)=\rho T(\bar u,s,u,x,t),
\end{veqnarray}
where $\rho $ is the number density of shocks, and $T(\bar u,s,u,x,t)$
is the PDF of
\begin{veqnarray}
  \nonumber (\bar u(y_0,t),s(y_0,t),u(y_0+x,t)),
\end{veqnarray}
conditional on $y_0$ being a shock position. Thus for statistically
homogeneous situations, (\ref{eq:3.15}) reduces to the following
equation for $Z(u_1,u_2,x,t)$
\begin{veqnarray}
  \label{eq:3.18}
  Z_t&=&-(u_2-u_1)Z_{x}\\
  &&- \int du'\ K(u_2-u')Z_x(u_1,u',x,t)\\
  &&+\int du'\ K(u_1-u')Z_x(u',u_2,x,t)\\
  &&+B_0 Z_{u_1u_1}+B_0 Z_{u_2u_2}+2B(x)Z_{u_1u_2}\\
  &&+G(u_1,u_2,x,t)+G(u_2,u_1,-x,t),
\end{veqnarray}
where
\begin{veqnarray}
  \label{eq:3.19}
  &&G(u_1,u_2,x,t)\\
  &=&\rho \biggl( \int_{-\infty}^0\!\! d s \ s
  \int_{u_1+s/2}^{u_1-s/2}\!\! d\bar u \ (u_1-\bar u)T(\bar
  u,s,u_2,x,t) \biggr)_{u_1}.
\end{veqnarray}

\subsection{Velocity difference and structure functions}
\label{sec:3.6}

Assume statistical homogeneity, and let $Z^\delta(\delta u,x,t)$ be
the PDF of the velocity difference $\delta u(x,y,t) =u(x+y,t)-u(y,t)$.
$Z^\delta(w,x,t)$ is related to $Z(u_1,u_2,x,t)$ by
\begin{veqnarray}
  \nonumber Z^\delta(w,x,t) =\int du\ 
  Z\left(u-\frac{w}{2},u+\frac{w}{2},x,t\right).
\end{veqnarray}
It then follows immediately from (\ref{eq:3.18}) and (\ref{eq:3.19})
that

\begin{corollary}
\label{th:3.4}

$Z^\delta$ satisfies
\begin{veqnarray}
  \label{eq:3.20}
  Z^\delta_t&=&-wZ^\delta_{x}-2\int dw'\ H(w'-w)
  Z^\delta_x(w',x,t)\\
  &&+2(B_0-B(x)) Z^\delta_{ww}+G^\delta(w,x,t),
\end{veqnarray}
where $K(w)=(H(w)-H(-w))/2$, $H(\cdot)$ is the Heaviside function and
\begin{veqnarray}
  \label{eq:3.21}
  G^\delta(w,x,t)&=&\int du\ 
  G\left(u-\frac{w}{2},u+\frac{w}{2},x,t\right)\\
  &+&\int du\ G\left(u+\frac{w}{2},u-\frac{w}{2},-x,t\right).
\end{veqnarray}

\end{corollary}

\demo{Remark} $G^\delta$ can be put into a form which is more convenient
for the calculations. Let
\begin{veqnarray}
  \nonumber
  \delta u_+(x,y_0,t)=u(y_0+|x|,t)-u_+(y_0,t),\\
  \delta u_-(x,y_0,t)= u_-(y_0,t)-u(y_0-|x|,t),
\end{veqnarray}
and let $U_\pm(s,\delta u_\pm,x,t)$ be the PDFs of $(s(y_0,t),\delta
u_\pm(x,y_0,t))$ conditional on $y$ being a shock position. Then,
$G^\delta$ can be expressed as
\begin{veqnarray}
  \nonumber G^\delta(w,x,t)=G^\delta_+(w,x,t)+G^\delta_-(w,x,t),
\end{veqnarray}
where
\begin{veqnarray}
  \nonumber G^\delta_\pm(w,x,t)&=&\frac{\rho }{2}\int_{-\infty}^0\!\!
  ds \ s\bigl(
  U_\pm(s,{\rm sgn}(x) w-s,x,t)+U_\pm(s,{\rm sgn}(x) w,x,t)\bigr)\\
  &-&\rho \int_{-\infty}^0\!\!  ds \ s\int_0^1d\beta \ U_\pm (s,{\rm
    sgn}(x) w-\beta s,x,t).
\end{veqnarray}
\enddemo

~

We now consider some consequences of (\ref{eq:3.20}). We have the
following two theorems:

\begin{theorem}
\label{th:3.5}
In the limit as $x\to0$,
\begin{veqnarray}
  \label{eq:3.22}
  &Z^\delta(w,x,t)=\delta(w)-|x|(\rho\delta(w)+\< \xi\>
  \delta^1(w)-\rho S(w,t))+o(x),\\[4pt]
  &{\d Z^\delta(w,x,t)= (1-\rho |x|)\frac{1}{x}
    Q\left(\frac{w}{x},t\right) +|x|\rho S(w,t)+o(x),}
\end{veqnarray}
where $\delta^1(w)=d\delta(w)/dw$, $o(x)$ must be interpreted in the
sense of weak convergence.  Here $Q(\xi,t)$ is the PDF of $\xi(x,t)$,
the regular part of the velocity gradient, i.e.
\begin{veqnarray}
  \nonumber u_x(x,t)= \xi(x,t)+\sum_j s(y_j,t) \delta (y-y_j),
\end{veqnarray}
and $S(s,t)$ is the PDF of $s(y_0,t)$ conditional on $y_0$ being a
shock location.

\end{theorem}

\begin{theorem}[Structure function scaling]
\label{th:3.6}

Let
\begin{veqnarray}
   \label{eq:3.aa}
   \< |\delta u|^a\>= \int dw \ |w|^a Z^\delta(w,x,t).
\end{veqnarray}
In the limit as $x\to0$,
\begin{veqnarray}
  \label{eq:3.aaa}
  \< |\delta u|^a\> = \left\{
    \begin{array}{ll}
      {\d |x|^{a} \< |\xi|^a\>+o(x^a)\quad}& \mbox{if}\ \ 0\le a<1,
      \\[6pt]
      {\d |x| \bigl( \< |\xi|\>+\rho \< |s|\>\bigr) 
        + o(x)\quad}&\mbox{if}\ \ a=1,\\[6pt]
      {\d |x| \rho \< |s|^a\> + o(x)\quad}&\mbox{if}\ \ 1 < a.
    \end{array}\right.
\end{veqnarray}

\end{theorem}

In terms of the moments, (\ref{eq:3.aaa}) is ($n\in\N_0$)
\begin{veqnarray}
  \label{eq:3.bbb}
  \< \delta u^{2n}\>&=&|x| \rho \< s^{2n}\> + o(x),\\
  \< \delta u^{2n+1}\>&=&x \rho \< s^{2n+1}\> + o(x).
\end{veqnarray}
The first moment satisfies $\< \delta u\> =0$ for all $(x,t)$.  This
is a consequence of statistical homogeneity and it is readily verified
since multiplying (\ref{eq:3.20}) by $w$ and integrating gives in $\<
\delta u\>_t=0$.

An equation for $Q(\xi,t)$ will be derived in Sec.~\ref{sec:4}.  To
interpret (\ref{eq:3.22}) note that $Z^\delta$ can be decomposed into
\begin{veqnarray}
  \nonumber
  Z^\delta(w,x,t)&=& p_{ns}(x,t) Z^\delta(w,x,t|\hbox{no shock})\\
  &+&(1-p_{ns}(x,t)) Z^\delta(w,x,t|\hbox{shock}),
\end{veqnarray}
where $p_{ns}(x,t)$ is the probability that there is no shock in
$[y,y+x)$, $Z^\delta(w,x,t|\hbox{no shock})$ is the PDF of $\delta
u(x,y,t)$ conditional on the property that there is no shock in
$[y,y+x)$, $Z^\delta(w,x,t|\hbox{shock})$ is the PDF of $\delta
u(x,y,t)$ conditional on the property that there is at least one shock
in $[y,y+x)$.  Since, by definition of~$\rho$ we have
\begin{veqnarray}
  \nonumber p_{ns} = 1 -\rho |x| + o(x),
\end{veqnarray}
(\ref{eq:3.22}) states that
\begin{veqnarray}
  \nonumber Z^\delta(w,x,t|\hbox{no shock})
  &=& (1-\rho |x|)\frac{1}{x} Q\left(\frac{w}{x},t\right)+o(x),\\
  Z^\delta(w,x,t|\hbox{shock})&=&S(w,t)+o(1).
\end{veqnarray}
This is consistent with the picture that $\delta u(x,y,t)=x
\xi(y,t)+o(x)$ if there is no shock in $[y,y+x)$, and
$w(x,y,t)=s(y_0,t)+o(1)$ if $y_0\in[y,x+y)$ is a shock position.

\begin{proof}[Proof of Theorem \ref{th:3.5}]
  
  We will consider the case $x>0$. The case $x<0$ can be treated
  similarly.  Note first that, in the limit as $x\to0$, we have
\begin{veqnarray}
  \nonumber x Z^\delta(x\xi,x,t)\to Q(\xi,t),
\end{veqnarray}
weakly. We postpone the proof of this fact until Sec.~\ref{sec:4}. It
implies that
\begin{veqnarray}
  \nonumber Z^\delta(w,x,t)= \delta(w)+o(1),
\end{veqnarray}
weakly. Define
\begin{veqnarray}
  \nonumber A(w,t)= \lim_{x\to0} x^{-1}(Z^\delta(w,x,t)-\delta(w))=
  \lim_{x\to0} {Z}_x^\delta(w,x,t).
\end{veqnarray}
Taking the limit as $x\to0$ in the equation for $Z^\delta$, it follows
that $A$ satisfies
\begin{veqnarray}
  \label{eq:3.a.1}
  0=-wA-2\int dw'\ H(w-w') A(w',x,t)+B(w,t),
\end{veqnarray}
where we used $\lim_{x\to0}(B_0-B(x))=0$ and we defined
\begin{veqnarray}
  \nonumber B(w,t)=\lim_{x\to0}G^\delta(w,x,t).
\end{veqnarray}
To evaluate $B$ note that as $x\to0$
\begin{veqnarray}
  \nonumber \delta u_\pm (x,y_0,t)\to0,
\end{veqnarray}
almost surely. This implies that, as $x\to0$,
\begin{veqnarray}
  \nonumber U_\pm(s,w,x,t) \to S(s,t)\delta(w),
\end{veqnarray}
where $S(s,t)$ is the PDF of $s(y_0,t)$ conditional on $y_0$ being a
shock location. Hence, from the expression for $G^\delta$,
\begin{veqnarray}
  \nonumber B(w,t)=\rho w S(w,t)+\rho \< s\> \delta(w) +2\rho
  \int_{-\infty}^w \!\! dw' S(w't)- 2\rho H(w),
\end{veqnarray}
where $H(\cdot)$ is the Heaviside function and we used $S(s,t)=0$ for
$s>0$ since $s(y_0,t)\le0$.  Inserting this expression in
(\ref{eq:3.a.1}), the solution of this equation is
\begin{veqnarray}
  \nonumber A(w,t)=-\delta(w)+\rho \< s\> \delta^1(w)+\rho S(w,t).
\end{veqnarray}
Here $\delta^1(w)=d\delta(w)/dw$ and we used the identity
$w\delta^1(w)=-\delta(w)$. Using (\ref{eq:4.3.1}), $\rho\< s\>=-\<
\xi\>$, this can be restated as
\begin{veqnarray}
  \nonumber A(w,t)=-\delta(w)-\< \xi\> \delta^1(w)+\rho S(w,t).
\end{veqnarray}
Hence, combining the above results, we have
\begin{veqnarray}
  \nonumber Z^\delta(w,x,t)=\delta(w)-x(\delta(w)+\< \xi\>
  \delta^1(w)- \rho S(w,t))+o(x),
\end{veqnarray}
weakly. This establishes the first equation in (\ref{eq:3.22}) for
$x>0$.  Reorganizing this expression as
\begin{veqnarray}
   Z^\delta(w,x,t) &=&(1-\rho x)(\delta(w)-x\< \xi\> \delta^1(w))+ x\rho
  S(w,t)+o(x),
\end{veqnarray}
weakly, and using the identity
\begin{veqnarray}
  \nonumber \delta(w)-x\< \xi\> \delta^1(w)=\frac{1}{x}
  Q\left(\frac{w}{x},t\right)+o(x),
\end{veqnarray}
establishes the second equation in (\ref{eq:3.22}) for $x>0$.

\end{proof}

\begin{proof}[Proof of Theorem \ref{th:3.6}]

  We will prove (\ref{eq:3.aaa}) directly for moments of integer order
  higher than one. For other values of $a$ (\ref{eq:3.aaa}) follows
  from (\ref{eq:3.22}) and the fact that the tails are controlled by
  higher order moments. Note first that for $a>0$
\begin{veqnarray}
  \nonumber \lim_{x\to0} \< |\delta u|^a\> =0,
\end{veqnarray}
since $\delta u(x,y,t)\to0$ almost surely as $x\to0$. Now, multiply
(\ref{eq:3.20}) by $w^{n}$ ($n\in \N$, $n\ge 2$), integrate and take
the limit as $x\to0\pm$. The result is
\begin{veqnarray}
  \nonumber 0=-a^\pm_{n+1}+\frac{2}{n+1} a^\pm_{n+1}+b^\pm_{n},
\end{veqnarray}
where
\begin{veqnarray}
  \nonumber a^\pm_n=\lim_{x\to 0\pm} x^{-1} \< \delta u^n\>
  =\lim_{x\to 0\pm}\< \delta u^n\>_x,
\end{veqnarray}
\begin{veqnarray}
  \nonumber b^\pm_n=\lim_{x\to 0\pm} \int dw \ w^n G^\delta(w,x,t).
\end{veqnarray}
Note that
\begin{veqnarray}
  \nonumber \int dw \ w^n G^\delta(w,x,t) &=& \frac{\rho}{2} \<
  s(\delta u_++{\rm sgn}(x) s)^n\>
  +\frac{\rho}{2} \< s\delta u_+^n\>\\
  &-&\rho \int_0^1 \!\! d\beta\ 
  \< s(\delta u_++\beta{\rm sgn}(x) s)^n\>\\
  &+&\frac{\rho}{2} \< s( \delta u_-+{\rm sgn}(x)s)^n\>
  +\frac{\rho}{2} \< s\delta u_-^n\>\\
  &-&\rho \int_0^1 \!\! d\beta\ \< s(\delta u_-+\beta{\rm sgn}(x)
  s)^n\>.
\end{veqnarray}
Since $\delta u_\pm (x,y_0,t)\to 0$ almost surely as $x\to0$, we have
\begin{veqnarray}
  \nonumber b^\pm_n= (\pm1)^{n} \rho \< s^{n+1}\>
  \left(1-\frac{2}{n+1}\right).
\end{veqnarray}
Inserting this expression in the equation for $a^\pm_n$ gives
\begin{veqnarray}
  \nonumber a^\pm_{n+1}=\lim_{x\to 0\pm} x^{-1} \< \delta u^{n+1}\>=
  (\pm1)^{n}\rho \< s^{n+1}\>.
\end{veqnarray}
Thus
\begin{veqnarray}
  \nonumber \< \delta u^{2n}\> = |x|\rho \< s^{2n}\>+o(x),\qquad \<
  \delta u^{2n+1}\> = x\rho \< s^{2n+1}\>+o(x).
\end{veqnarray}
This proves (\ref{eq:3.bbb}).

We now prove (\ref{eq:3.aaa}) for $0\le a\le1$.  The proof for other values
of $a$ is similar.  Let
\begin{veqnarray}
  \nonumber f^\delta(w,x,t)&=&(1-\rho |x|)\frac{1}{x}
  Q\left(\frac{w}{x},t\right)
  +|x|\rho S(w,t),\\[4pt]
  g^\delta(w,x,t)&=&\delta(w)-|x|(\rho \delta(w)+\< \xi\>
  \delta^1(w)-\rho S(w,t)),
\end{veqnarray}
and write for $M>0$
\begin{veqnarray}
  \nonumber
  &&\int dw \ |w|^a (Z^\delta-f^\delta)\\
  &=& \int_{|w|\le M}\!\! dw \ |w|^a (Z^\delta-f^\delta)+ \int_{|w|>
    M}\!\! dw \ |w|^a (Z^\delta-f^\delta).
\end{veqnarray}
The first term at the right hand-side is $o(x)$ because of
(\ref{eq:3.22}). To estimate the second term, note that for $M$ large
enough
\begin{veqnarray}
  \nonumber \int_{|w|> M}\!\! dw \ |w|^a Z^\delta&\le&
  \int_{|w|> M}\!\! dw \ w^2 Z^\delta\\
  &\le& \left|\int dw \ w^2 (Z^\delta-g^\delta)\right|+
  \int_{|w|> M}\!\! dw \ w^2 g^\delta\\
  &=&o(x) +|x|\rho \int_{|w|> M}\!\! dw \ w^2 S(w,t)\\
  &=&o(x)+O(x)\int_{|w|> M}\!\! dw \ w^2 S(w,t).
\end{veqnarray}
\begin{veqnarray}
  \nonumber \int_{|w|> M}\!\! dw \ |w|^a f^\delta&=& |x|^a (1-|x|\rho)
  \int_{|\xi|> M/x}\!\! d\xi \ |\xi|^a Q(\xi,t) \\
  &&+ |x|\rho\int_{w> M}\!\! dw \ |w|^a S(w,t)\\
  &=&o(x^a)+O(x)\int_{w> M}\!\! dw \ |w|^a S(w,t).
\end{veqnarray}
Since $M$ can be made arbitrarily large, we get
\begin{veqnarray}
  \nonumber \int dw \ |w|^a (Z^\delta-f^\delta)\le o(x^a)+\delta_M O(x),
\end{veqnarray}
where $\delta_M\to0$ as $M\to+\infty$. Noting that
\begin{veqnarray}
  \nonumber 
  \int dw \ |w|^a f^\delta= \left\{
    \begin{array}{ll}
      {\d |x|^{a} \< |\xi|^a\>\quad}& \mbox{if}\ \ 0\le a<1,
      \\[6pt]
      {\d |x| \bigl( \< |\xi|\>+\rho \< |s|\>\bigr) 
        \quad}&\mbox{if}\ \ a=1,
    \end{array}\right.
\end{veqnarray}
we obtain (\ref{eq:3.aaa}) for $0\le a\le 1$.

\end{proof}

\section{Velocity gradient}
\label{sec:4}

We now turn to the study of the PDF of the velocity gradient for the
solution of (\ref{eq:3.3}). Since the solutions typically contain
discontinuities, it is already an issue whether the PDF for the
velocity gradient is well-defined. Heuristically, this is so since $u$
only fails to be differentiable at no more than countably many points.
We will therefore be concerned with the PDF of the regular part of the
gradient.

\subsection{Master equation for the PDF of the velocity gradient}
\label{sec:4.1}

We focus on the statistically homogeneous case and derive an equation
for $Q(\xi,t)$, the PDF of $\xi(x,t)$, defined as the regular part of
the velocity gradient, i.e.
\begin{veqnarray}
  \nonumber u_x(x,t)=\xi(x,t)+\sum_j s(y_j) \delta(x-y_j),
\end{veqnarray}
where $\xi(\cdot,t)\in L^1(I)$.  We will prove that

\begin{theorem}
\label{th:5}

$Q$ satisfies
\begin{veqnarray}
  \label{eq:4.1}
  Q_t=\xi Q +\bigl(\xi^2 Q\bigr)_\xi+ B_1Q_{\xi\xi}+F(\xi,t),
\end{veqnarray}
where $B_1=-B_{xx}(0)$ and
\begin{veqnarray}
  \label{eq:4.2}
  F(\xi,t)=\rho\int_{-\infty}^0\!\! ds \ s V(s,\xi,t).
\end{veqnarray}
Here $V(s,\xi,t)=(V_+(s,\xi,t)+V_-(s,\xi,t))/2$, $V_\pm(s,\xi_\pm,t)$
are the PDFs of
\begin{veqnarray}
  \nonumber (s(y_0,t),\xi_\pm(y_0,t)=u_x(y_0\pm,t)),
\end{veqnarray}
conditional on the property that $y_0$ is a shock position.

\end{theorem}

The consequences of (\ref{eq:4.1}) will be studied in
Sec.~\ref{sec:4.3}.  Note that if $u(x,t)\law -u(-x,t)$, then
\begin{veqnarray}
  \nonumber (\xi_+(x,t),s(x,t))\law(\xi_-(-x,t),s(-x,t)),
\end{veqnarray}
and $V_+(s,\xi,t)=V_-(s,\xi,t)=V(s,\xi,t)$.

\begin{proof}[Proof of Theorem \ref{th:5}]
  Let $Q^\delta (\eta,x,t)$ be the PDF of
  $\eta(x,y,t)=(u(x+y,t)-u(y,t))/x$.  $Q^\delta$ is related to $Z$ and
  $Z^\delta$ by
\begin{veqnarray}
  \nonumber Q^\delta (\eta,x,t)=x\, Z^\delta(x\eta,x,t) = x \int du \ 
  Z\left(u-\frac{x\eta}{2},u+\frac{x\eta}{2},x,t\right).
\end{veqnarray}
From (\ref{eq:3.20}) it follows that $Q^\delta $ satisfies
\begin{veqnarray}
  \label{eq:4.3}
  Q^\delta_t&=&\eta Q^\delta+\bigl(\eta^2 Q^\delta \bigr)_\eta+
  B_1(x)Q^\delta_{\eta\eta}
  -x\eta Q^\delta_x\\
  && -2x\int d\eta'\ H(\eta-\eta') Q^\delta_x(\eta',x,t)
  +F^\delta(\eta,x,t),
\end{veqnarray}
where $B_1(x)=2(B_0-B(x))/x^2$ and
\begin{veqnarray}
  \nonumber F^\delta(\eta,x,t)=
  F^\delta_1(\eta,x,t)+F^\delta_2(\eta,x,t)+F^\delta_3(\eta,x,t),
\end{veqnarray}
with
\begin{veqnarray}
  \nonumber F^\delta_1(\eta,x,t)&=& \rho\int_{-\infty}^0\!\! ds \ s
  V^\delta \left(s,\eta,x,t\right),\\
  F^\delta_2(\eta,x,t)&=&\rho\int_{-\infty}^0\!\! ds \ s
  V^\delta\left(s,\eta-\frac{s}{|x|},x,t\right),\\
  F^\delta_3(\eta,x,t)&=&-2\rho \int_{-\infty}^0\!\! ds \ 
  s\int_0^1d\beta\ V^\delta \left(s,\eta-\frac{\beta
      s}{|x|},x,t\right).
\end{veqnarray}
Here
$V^\delta(s,\eta,x,t)=(V^\delta_+(s,\eta,x,t)+V^\delta_-(s,\eta,x,t))/2$,
$V^\delta_\pm(s,\eta,x,t)$ are the PDFs of
\begin{veqnarray}
  \nonumber (s(y_0,t),\eta_\pm(y_0,x,t)),
\end{veqnarray}
with $\eta_\pm(y_0,x,t)=\pm(u(y_0\pm|x|,t)-u_\pm(y_0,t))/|x|$,
conditional on the property that $y_0$ is a shock position.

Define
\begin{veqnarray}
  \nonumber Q(\xi,t)=\lim_{x\to0}Q^\delta(\xi,x,t).
\end{veqnarray}
It is easy to see that
\begin{veqnarray}
  \nonumber
  &&\int d\eta\  F^\delta_1(\eta,x,t) = \rho\< s\>,\\
  &&\int d\eta \ F^\delta_2(\eta,x,t) = \rho\int d\eta
  \int_{-\infty}^0\!\! ds \ s
  V^\delta(s,\eta,x,t)=\rho\< s\>,\\
  &&\int d\eta\ F^\delta_3(\eta,x,t) = -2\rho\int d\eta
  \int_{-\infty}^0\!\!  ds \ s V^\delta(s,\eta,x,t)=-2\rho\< s\>,
\end{veqnarray}
consistent with the fact that
\begin{veqnarray}
  \nonumber \int d\eta \ F^\delta(\eta,x,t) =0,
\end{veqnarray}
and, hence,
\begin{veqnarray}
  \nonumber \frac{d}{dt}\int d\eta\ Q^\delta(\eta,x,t)=0, \qquad \int
  d\eta\ \eta Q^\delta(\eta,x,t)=0.
\end{veqnarray}
In the limit as $x\to0$, $\eta_+$ and $\eta_-$ converge respectively
to the gradient of the velocity at the left and right sides of the
shock, and, pointwise in $\eta$, we have
\begin{veqnarray}
  \nonumber \lim_{x\to0}F^\delta_2 (\eta,x,t)=\lim_{x\to0}F^\delta_3
  (\eta,x,t)=0,
\end{veqnarray}
\begin{veqnarray}
  \nonumber \lim_{x\to0}F^\delta_1 (\eta,x,t)=F (\eta,t),
\end{veqnarray}
with $F$ given by (\ref{eq:4.2}). Therefore, a standard argument with
test functions applied to (\ref{eq:4.3}) shows that in the limit as
$x\to0$, $Q^\delta$ converges weakly to $Q$, solution of
(\ref{eq:4.1}).

\end{proof}

\demo{Remark} $F^\delta_2$ and $F^\delta_3$ are examples of what we
will call ``ghost terms'', i.e. terms that have finite total moments,
but in the limit converge pointwise to zero. Due to the ghost terms,
is not clear at the moment that $Q$ satisfies $\int d\xi
\,Q(\xi,t)=1$. This will be established as a consequence of
Lemma~\ref{lem:1}. \enddemo

\subsection{Alternative limiting processes}
\label{sec:4.2}

Using BV-calculus, one works at $\nu=0$ and access the statistics of
the velocity gradient by taking $x\to0$ in $(u(x+y,t)-u(y,t))/x$.
This procedure gives an equation for
\begin{veqnarray}
  \nonumber Q(\xi,t)=\lim_{x\to0} Q^\delta(\xi,x,t),
\end{veqnarray}
where $Q^\delta(\xi,x,t)$ is the PDF of $(u(x+y,t)-u(y,t))/x$.  In
this section, we revert the order of the limits: we take $x\to0$
first, working at finite $\nu$, and then let $\nu\to0$. As in
Sec.~\ref{sec:3.2}, this is done using boundary layer analysis and
matched asymptotics. In this way, we will obtain an equation for
\begin{veqnarray}
  \nonumber \underline{Q}(\xi,t)=\lim_{\nu\to0} Q^\nu(\xi,t),
\end{veqnarray}
which will turn out to be identical to the one for $Q$. To the extend
that boundary layer analysis can be justified, this strongly suggests
that the limits $x\to0$, $\nu\to0$ commute.

For statistically homogeneous situations, recall that $Q^\nu(\xi,t)$
satisfies (see (\ref{eq:A.6}) in the Appendix)
\begin{veqnarray}
  \nonumber Q^\nu_t=\xi Q^\nu +\bigl(\xi^2 Q^\nu\bigr)_\xi +B_1
  Q^\nu_{\xi\xi} -\nu\bigl( \< \xi_{xx}|\xi\> Q^\nu\bigr)_\xi.
\end{veqnarray}
The average of the dissipative term can be expressed as
\begin{veqnarray}
  \nonumber \nu \< \xi_{xx}|\xi\> Q^\nu =
  \nu\lim_{L\to\infty}\frac{1}{2L} \int_{-L}^{L} dx\ \xi_{xx}(x,t)
  \delta(\xi-\xi(x,t)).
\end{veqnarray}
As in Sec.~\ref{sec:3.2}, we will evaluate this integral using the
approximation for $\xi_{xx}$ provided by the boundary layer analysis.
However, this analysis has to be developed further in order to
evaluate the dissipative term for the velocity gradient. Recall that
$\xi^\inn = u^\inn_x = {v_0}_z/\nu+{v_1}_z+O(\nu)$.  Inside the shock,
the $O(1)$ contribution of ${v_1}_z$ is clearly negligible compared to
the $O(\nu^{-1})$ contribution of ${v_0}_z/\nu$. However, the
contribution of ${v_1}_z$ {\em is} important at the border of the
shock because ${v_0}_z$ decays exponentially fast there.

To evaluate ${v_1}_z$ at the shock boundaries $z\to\pm\infty$,
consider the equation for $v_1$ that we get from (\ref{eq:3.13}):
\begin{veqnarray}
  \nonumber {v_0}_t+(v_0-\bar u){v_1}_z+{v_0}_z(v_1+\gamma) =
  {v_1}_{zz}+f.
\end{veqnarray}
The general solution of this equation can be expressed as
\begin{veqnarray}
  \nonumber
  v_1 &=& C_1 {\rm e}^{\int_0^z\! dz'\, (v_0(z')-\bar u)}\\
  &+& \int_0^z\!\!dz' \biggl(C_2+(v_0(z')-\bar u)\gamma+
  \int_0^{z'}dz'' {v_0}_t(z'')-f z'\biggr) {\rm e}^{\int_{z'}^z\!
    dz''\, (v_0(z'')-\bar u)},
\end{veqnarray}
where $C_1$, $C_2$ are constants. They, as well as $\gamma$, have to
be determined by matching with $u^\out=u_0+u_1+O(\nu^2)$
\cite{goxi92}. We will not dwell on this problem since $C_1$, $C_2$,
$\gamma$ do not enter the expression for ${v_1}_z$ as $z\to\pm\infty$.
Indeed, using the expression for $v_0$, direct computation shows that
$v_1$ reduces asymptotically to
\begin{veqnarray}
  \nonumber v_1 &=& \frac{1}{s}
  \biggl(2\frac{d\bar u}{dt}-\frac{ds}{dt} -2f\biggr) z\\
  &+&\frac{2}{s^2}\biggl(C_2s+\frac{ds}{dt} -2\frac{d\bar
    u}{dt}-\frac{s^2\gamma}{2}+2f\biggr) +O\left({\rm
      e}^{-sz/2}\right)\qquad \hbox{as } z\to-\infty,
\end{veqnarray}
and
\begin{veqnarray}
  \nonumber v_1&=& -\frac{1}{s}
  \biggl(2\frac{d\bar u}{dt}+\frac{ds}{dt} -2f\biggr) z\\
  &-&\frac{2}{s^2}\biggl(C_2s+\frac{ds}{dt}+2\frac{d\bar u}{dt}
  +\frac{s^2\gamma}{2}-2f\biggr)+O\left({\rm e}^{sz/2}\right)\qquad
  \hbox{as } z\to+\infty.
\end{veqnarray}
Thus
\begin{veqnarray}
  \nonumber \lim_{z\to\pm\infty} {v_1}_z = \mp \frac{2}{s}\frac{d\bar
    u}{dt} -\frac{1}{s} \frac{ds}{dt} \pm\frac{2f}{s}\equiv\xi_\pm,
\end{veqnarray}
where the last equality is simply a definition of $\xi_\pm$. Note that
these can be reorganized to give
\begin{veqnarray}
  \nonumber \frac{ds}{dt} = -\frac{s}{2} \bigl( \xi_+ +\xi_-\bigr),
  \qquad \frac{d\bar u}{dt} = -\frac{s}{4} \bigl( \xi_+
  -\xi_-\bigr)+f.
\end{veqnarray}
In the limit as $\nu\to0$ these are the equations of motion along the
shock.

Using these results, to $O(\nu)$, (\ref{eq:3.12}) can be estimated as
\begin{veqnarray}
  \nonumber \nu \< \xi_{xx} |\xi\> Q^\nu &=& \nu \rho \int ds d\bar
  ud\xi_+ d\xi_- \ S
  (\bar u,s,\xi_+,\xi_-,t) \\
  &&\qquad \times \int_{\Omega} dx\ \xi^\inn_{xx}(x,t)
  \delta(\xi-\xi^\inn(x,t)),
\end{veqnarray}
where $\Omega$ is a layer centered at $y$ with width $\gg O(\nu)$,
$S(\bar u,s,\xi_+,\xi_-,t)$ is the PDF of $(\bar
u(y_0,t),s(y_0,t),\xi_+(y_0,t),\xi_-(y_0,t))$, conditional on $y_0$
being a shock location ($\xi_+$ and $\xi_-$ have to be included for
reasons to be made clear later).

We now go to the stretched coordinate $z=(x-y)/\nu$, and use the
result of Sec.~\ref{sec:3.2}, namely,
\begin{veqnarray}
  \nonumber
  \label{eq:xiin}
  \xi^\inn = \nu^{-1} {v_0}_z +{v_1}_z +O(\nu).
\end{veqnarray}
To $O(\nu)$, both terms must be included. However the contribution of
${v_1}_z$ is important only at the border of the shock where ${v_0}_z$
falls exponentially fast, and can be neglected inside the shock.
Thus, $\xi^\inn \approx \bar \xi= \nu^{-1} {v_0}_z +\xi_\pm$, and we
have
\begin{veqnarray}
  \nonumber \nu \< \xi_{xx} |\xi\> Q^\nu &=&\rho \int ds d\bar ud\xi_+
  d\xi_- \ S
  (\bar u,s,\xi_+,\xi_-,t)\\
  &&\qquad \times\int_{-\infty}^{+\infty}\!\! dz \ \bar\xi_{zz}(z,t)
  \delta(\xi-\bar\xi(z,t)).
\end{veqnarray}
To perform the $z$ integral, we change the integration variable to
$\xi'=\bar \xi-\xi_-$ for $z<0$ and to $\xi'=\bar \xi-\xi_+$ for
$z>0$.  For $z<0$
\begin{veqnarray}
  \nonumber
  &&((v_0-\bar u)(\bar \xi-\xi_-))_z=\bar \xi_{zz},\\
  &&(v_0-\bar u)=\left({\t\frac{1}{4}}s^2+2\nu(\bar
    \xi-\xi_-)\right)^{1/2}= -\frac{s}{2} +O(\nu),
\end{veqnarray}
and we have
\begin{veqnarray}
  \nonumber dz {\bar \xi}_{zz}=d\xi'\ \frac{\bar \xi_{zz}}{(\bar
    \xi-\xi_\pm)_z}=- d\xi'\ \frac{(s/2(\bar \xi-\xi_\pm))_z}{(\bar
    \xi-\xi_\pm) _z} = -d\xi' \frac{s}{2}.
\end{veqnarray}
Similarly, for $z>0$
\begin{veqnarray}
  \nonumber
  &&((v_0-\bar u)(\bar \xi-\xi_+))_z=\bar \xi_{zz},\\
  &&(v_0-\bar u)=-\left({\t\frac{1}{4}}s^2+2\nu(\bar
    \xi-\xi_+)\right)^{1/2}= \frac{s}{2} +O(\nu),
\end{veqnarray}
and we have
\begin{veqnarray}
  \nonumber dz {\bar \xi}_{zz}=d\xi'\ \frac{\bar \xi_{zz}}{(\bar
    \xi-\xi_\pm)_z}= d\xi'\ \frac{(s/2(\bar \xi-\xi_\pm))_z}{(\bar
    \xi-\xi_\pm) _z} = d\xi' \frac{s}{2}.
\end{veqnarray}
This leads to
\begin{veqnarray}
  \nonumber
  &&\nu \< \xi_{xx}|\xi\> Q^\nu\\
  &=&\rho \int ds d\bar ud\xi_+ d\xi_- \ S (\bar
  u,s,\xi_+,\xi_-,t)\int^{0}_{-s^2/(8\nu)}
  \!\! d \xi'\ \frac{s}{2}\, \delta(\xi-\xi'-\xi_-)\\
  &\ns{+}&\rho \int ds d\bar ud\xi_+ d\xi_- \ S (\bar
  u,s,\xi_+,\xi_-,t)\int_{-s^2/(8\nu)}^{0} \!\! d\xi'\ \frac{s}{2}\,
  \delta(\xi-\xi'-\xi_+).
\end{veqnarray}
Letting $\nu\to0$ and integrating gives
\begin{veqnarray}
  \nonumber \lim_{\nu\to0}\nu \< \xi_{xx}|\xi\> Q^\nu=
  \rho\int_{-\infty}^0\!\! ds\ s \int_\xi^{\infty}\!\! d\xi'\ 
  V(s,\xi',t),
\end{veqnarray}
where we used the consistency constraint
\begin{veqnarray}
  \nonumber \int d\bar u d\xi_\mp\, S (\bar u,s,\xi_+,\xi_-,t)
  =V_\pm(s,\xi_\pm,t).
\end{veqnarray}
Thus
\begin{veqnarray}
  \nonumber -\lim_{\nu\to0}\nu \bigl(\< \xi_{xx}|\xi\>
  Q^\nu\bigr)_\xi= \rho\int_{-\infty}^0\!\! ds\ s V(s,\xi,t)=F(\xi,t).
\end{veqnarray}

\subsection{Consequences of the master equation}
\label{sec:4.3}

First we observe
\begin{lemma}
\label{lem:1}
\begin{veqnarray}
  \label{eq:4.3.1}
  \< \xi\> + \rho \< s\>=0, \quad \hbox{or }\quad \int d\xi \ \xi
  Q(\xi,t) =-\rho \< s\>.
\end{veqnarray}
\end{lemma}

\begin{proof}
  Denote by $u^\nu$ the solution of (\ref{eq:1.1}) for finite $\nu$.
  Then $\< u^\nu_x\>=0$. Let $\varphi(x)$ be a compactly supported
  smooth function. We have
\begin{veqnarray}
  \nonumber 0=\int dx \ \varphi \< u^\nu_x\> =\Bigl\< \int dx \ 
  \varphi u^\nu_x\Bigr\> =-\Bigl\< \int dx \ \varphi_x u^\nu\Bigr\>.
\end{veqnarray}
In the limit as $\nu\to0$, $u^\nu\to u$, the solution of
(\ref{eq:3.3}).  Thus
\begin{veqnarray}
  \nonumber 0= -\Bigl\< \int dx \ \varphi_x u\Bigr\>.
\end{veqnarray}
Denote by $y_j$ the location of the shocks. We can write
\begin{veqnarray}
  \nonumber \int dx \ \varphi_x u=-\sum_j \int_{y_j}^{y_{j+1}}\!\! dx
  \ \varphi u_x-\sum_j \varphi(y_j) s(y_j,t).
\end{veqnarray}
Averaging this result, we get
\begin{veqnarray}
  \nonumber \int dx \ \varphi \bigl(\< \xi\> +\rho\< s\>\bigr) = 0,
\end{veqnarray}
for all compactly supported smooth functions $\varphi$. Hence
(\ref{eq:4.3.1}).

\end{proof}

Notice that for finite $\nu$ or $x$, $\int d\xi\, \xi Q^\nu(\xi,t)=
\int d\xi\, \xi Q^\delta(\xi,x,t)=0$. In the language of Kraichnan
\cite{kra99}, (\ref{eq:4.3.1}) represents a flow of probability of
$\xi$ from the smooth part of the velocity field to the shocks. It
reflects the fact that no matter how small $\nu$ is, the dissipation
range has a finite effect on inertial range statistics.

As a consequence of (\ref{eq:4.1}) and (\ref{eq:4.3.1}), we have
\begin{veqnarray}
  \nonumber \frac{d}{dt} \int d\xi \ Q(\xi,t)= 0,
\end{veqnarray}
i.e. the normalization of $Q$ is preserved. For the initial data we
are interested in, this implies
\begin{veqnarray}
  \nonumber \int d\xi \ Q(\xi,t)= 1.
\end{veqnarray}

\begin{lemma}
\label{lem:2}
\begin{veqnarray}
  \label{eq:4.5}
  \frac{d}{dt}(\rho\< s\>)=- \frac{\rho}{2}(\< s\xi_+\>+\< s\xi_-\>).
\end{veqnarray}
In particular $\int d\xi \, \xi F(\xi,t) = \rho(\< s\xi_+\>+\<
s\xi_-\>)/2$ exists and is finite.
\end{lemma}

\begin{proof}
  Since $\rho\< s\>$ and its derivative are obviously finite, the
  existence of the integral $\int d\xi \, \xi F(\xi,t) = \rho(\<
  s\xi_+\>+\< s\xi_-\>)/2$ follows from the argument below and a
  standard cut-off argument on large values of $\xi_\pm$.
  
  Using
\begin{veqnarray}
  \nonumber {u_\pm}_t+u_\pm \xi_\pm=f,
\end{veqnarray} 
where $\xi_\pm(x,t)=u_x(x\pm,t)$, and $dy_j/dt=\bar u(y_j,t)$, we have
 \begin{veqnarray}
   \nonumber
   \frac{d}{dt} u_+(y_j,t)&=& \frac{dy_j}{dt} \xi_+(y_j,t) +{u_+}_t(y_j,t)\\
   &=& \bar u(y_j,t)\xi_+(y_j,t) - u_+(y_j,t)\xi_+(y_j,t)+f(y_j,t)\\
   &=& -\frac{1}{2}s(y_j,t)\xi_+(y_j,t) + f(y_j,t).
 \end{veqnarray}
 Similarly
 \begin{veqnarray}
   \nonumber \frac{d}{dt} u_-(y_j,t)=\frac{1}{2}s(y_j,t)\xi_-(y_j,t) +
   f(y_j,t).
 \end{veqnarray}
 These equations can be reorganized into
 \begin{veqnarray}
   \nonumber \frac{d}{dt} \bar u(y_j,t)=
   -\frac{1}{4}s(y_j,t)(\xi_+(y_j,t)-\xi_-(y_j,t))+f(y_j,t).
 \end{veqnarray}
 and
 \begin{veqnarray}
   \nonumber \frac{d}{dt} s(y_j,t)=-
   \frac{1}{2}s(y_j,t)(\xi_+(y_j,t)+\xi_-(y_j,t)).
 \end{veqnarray}
 Consider now
\begin{veqnarray}
  \label{eq:4.6}
  \rho\< s\> = \biggl\< \sum_j s(x,t) \delta(x-y_j)\biggr\>= \biggl\<
  \sum_j s(y_j,t) \delta(x-y_j)\biggr\>.
\end{veqnarray}
Using the equation for $s(y_j,t)$ and $dy_j/dt=\bar u(y_j,t)$, we have
\begin{veqnarray}
  \nonumber \frac{d}{dt} (\rho\< s\>)&=&-
  \frac{\rho}{2}(\< s\xi_+\>+\< s\xi_-\>)\\
  &&-\biggl\< \sum_j
  s(y_j,t)\bar u(y_j,t) \delta(x-y_j)\biggr\>_x\\
  && +\hbox{contribution from shock creation and collision}.
\end{veqnarray}
The second term at the right hand-side is zero by homogeneity. Thus to
obtain (\ref{eq:4.5}) it remains to prove that the contribution of
shock creation and collision vanish. To understand how this term
arises and why it vanishes, assume a shock is created at position
$y_1$ at time $t_1$. Then the sum under the average in (\ref{eq:4.6})
involves a term like
\begin{veqnarray}
  \nonumber T_1=s(y_1,t)\delta(x-y_1) H(t-t_1).
\end{veqnarray}
where $H(\cdot)$ is the Heaviside function.  Time differentiation
gives
\begin{veqnarray}
  \nonumber \frac{dT_1}{dt}&=&\frac{d}{dt}s(y_1,t) \delta(x-y_1)
  H(t-t_1)-
  s(y_1,t)\bar u(y_1,t)\delta_x(x-y_1) \\
  &+& s(y_1,t)\delta(x-y_1) \delta(t-t_1).
\end{veqnarray}
The last term accounts for shock creation.  Since the shock amplitude
is zero at creation,
\begin{veqnarray}
  \nonumber s(y_1,t)\delta(x-y_1) \delta(t-t_1)
  =s(y_1,t_1)\delta(x-y_1) \delta(t-t_1)=0.
\end{veqnarray}
This means that shock creation makes no contribution to the time
derivative of (\ref{eq:4.6}).  Consider now the collision events.
Assume at time $t_1$ the shocks located at $y_2$ and $y_3$ merge into
one shock located at $y_1$. Obviously $y_1(t_1)=y_2(t_1)=y_3(t_1)$.
Such an event contributes to the sum under the average in
(\ref{eq:4.6}) by a term like
\begin{veqnarray}
  \nonumber
  T_2&=&s(y_1,t)\delta(x-y_1) H(t-t_1)\\
  &+&\bigl(s(y_2,t)\delta(x-y_2)+s(y_3,t)\delta(x-y_3)\bigr)H(t_1-t).
\end{veqnarray}
Time differentiation gives
\begin{veqnarray}
  \nonumber \frac{dT_2}{dt}&=&
  \frac{d}{dt}s(y_1,t)\delta(x-y_1) H(t-t_1)\\
  &+&\frac{d}{dt}s(y_2,t)\delta(x-y_2) H(t_1-t)
  +\frac{d}{dt}s(y_3,t)\delta(x-y_3)H(t_1-t)\\
  &-& s(y_1,t)\bar u(y_1,t) \delta_x(x-y_1) H(t-t_1)\\
  &-&\bigl(s(y_2,t)\bar u(y_2,t)\delta_x(x-y_2)
  +s(y_3,t)\bar u(y_3,t)\delta_x(x-y_3)\bigr)H(t_1-t)\\
  &+& \bigl(s(y_1,t)\delta(x-y_1)
  -s(y_2,t)\delta(x-y_2)-s(y_3,t)\delta(x-y_3)\bigr) \delta(t-t_1).
\end{veqnarray}
The term involving $\delta(t-t_1)$ arises from shock collision.  Since
$y_1(t_1)=y_2(t_1)=y_3(t_1)$ and shock amplitudes add up at collision,
\begin{veqnarray}
  \nonumber
  &&\lim_{t\to0+}s(y_1(t_1+t),t_1+t)\delta(x-y_1(t_1+t))\\
  &=&\lim_{t\to0+}
  \bigl(s(y_2(t_1-t),t_1-t)\delta(x-y_2(t_1-t))\\
  &&\quad\ \, +s(y_3(t_1-t),t_1-t) \delta(x-y_3(t_1-t))\bigr),
\end{veqnarray}
the term in the equation for $T_2$ involving $\delta(t-t_1)$ vanishes.
This means shock collision makes no contribution to the time
derivative of (\ref{eq:4.6}). Hence (\ref{eq:4.5}).
\end{proof}

\begin{corollary}
\label{lem:4}
We have
\begin{veqnarray}
  \label{eq:4.8}
  \lim_{|\xi|\to\infty} |\xi|^3 Q(\xi,t)=0,
\end{veqnarray}
i.e. $Q$ decays faster than $|\xi|^{-3}$ as $\xi\to-\infty$.

\end{corollary}

\begin{proof}
  Taking the first moment of (\ref{eq:4.1}) leads to
\begin{veqnarray}
  \nonumber \frac{d}{dt}\< \xi\>
  =\left[\xi^3Q\right]_{-\infty}^{+\infty} + \int d\xi \ \xi
  F=\left[\xi^3Q\right]_{-\infty}^{+\infty} + \frac{\rho}{2} \left(\<
    s\xi_+\>+\< s\xi_-\>\right)
\end{veqnarray}
Using (\ref{eq:4.3.1}) this equation can be written as
\begin{veqnarray}
  \nonumber \frac{d}{dt}\bigl(\rho\< s\> \bigr)
  =-\left[\xi^3Q\right]_{-\infty}^{+\infty} - \frac{\rho}{2} \left(\<
    s\xi_+\>+\< s\xi_-\>\right).
\end{veqnarray}
Using (\ref{eq:4.5}) we obtain that the boundary term must be zero.
Since $\xi^3Q$ has different sign for large negative and positive
values of $\xi$, one must have separately
$\lim_{\xi\to-\infty}|\xi|^3Q=0$ and $\lim_{\xi\to+\infty}|\xi|^3Q=0$.
Hence (\ref{eq:4.8}).
\end{proof}

\demo{Remark} Lemma \ref{lem:2} uses the fact that shocks are created
at zero amplitude and shock strengths add up at collision, which holds
in the case of large-scale smooth forcing \cite{ekhma99}. It is
possible to construct pathological situations, such as the unforced
case with piecewise linear initial data \cite{pol99}, in which case
shocks are created at finite amplitude. In this case Lemma \ref{lem:2} is
changed to

\begin{lemma}
\label{lem:3}

\begin{veqnarray}
  \label{eq:4.7}
  \frac{d}{dt}(\rho\< s\>)=- \frac{\rho}{2}(\< s\xi_+\>+\<
  s\xi_-\>)-D_c,
\end{veqnarray}
where
\begin{veqnarray}
  \nonumber D_c=-\sigma_1 \< s_1\>-\sigma_2 \< s_2\>\ge0.
\end{veqnarray}
Here $\sigma_1$ and $\sigma_2$ are respectively the space-time number
density of shock creation and collision points, $\< s_1\>\le0$ is the
average shock amplitude at creation, and $\< s_2\>\le0$ is the average
gain of amplitude at shock collision. In this case
\begin{veqnarray}
  \label{eq:4.8.1}
  Q(\xi,t)\sim D_c\, |\xi|^{-3}\qquad \hbox{as } \xi\to-\infty.
\end{veqnarray}
\end{lemma}

Lemma \ref{lem:3} follows from a direct adaptation of the proof of
Lemma \ref{lem:2}.
\enddemo

\subsection{Realizability and asymptotics for the statistical steady state}
\label{sec:4.4}

We now turn to the study of (\ref{eq:4.1}) at steady state
\begin{veqnarray}
  \label{eq:4.4.1}
  0=\xi Q +\bigl(\xi^2 Q\bigr)_\xi+ B_1 Q_{\xi\xi}+F(\xi),
\end{veqnarray} 
where $F(\xi)=\lim_{t\to\infty} F(\xi,t)$. We shall prove that

\begin{theorem}
  \label{th:4.6}
  The realizability constraint $Q\in L^1(\R)$, $Q\ge0$ implies
\begin{veqnarray}
  \label{eq:4.4.2}
  \lim_{\xi\to+\infty} \xi^{-2} {\rm e}^{\Lambda} F(\xi)=0,
\end{veqnarray}
where $\Lambda=-\xi^3/3B_1$. Assuming (\ref{eq:4.4.2}), the only
positive solution of (\ref{eq:4.4.1}) can be expressed as
\begin{veqnarray}
  \label{eq:4.4.3}
  Q_\infty(\xi)= \frac{1}{B_1}\int_{-\infty}^\xi \!\! d\xi' \ \xi'
  F(\xi')-\frac{\xi{\rm e}^{-\Lambda}}{B_1} \int_{-\infty}^\xi \!\!
  d\xi' \ {\rm e}^{\Lambda'} G(\xi'),
\end{veqnarray}
where
\begin{veqnarray}
  \nonumber G(\xi)=F(\xi)+\frac{\xi}{B_1}
  \int_{-\infty}^{\xi}\!\!d\xi'\ \xi' F(\xi').
\end{veqnarray}
Furthermore
\begin{veqnarray}
  \label{eq:4.4.4}
  Q_\infty(\xi)\sim \left\{
    \begin{array}{ll}
      {\d |\xi|^{-3} \int_{-\infty}^\xi\!\! d\xi' \xi' F(\xi') 
        \qquad}&\hbox{as  } \xi\to-\infty,\\
      {\d C_+ \xi{\rm e}^{-\Lambda}}
        &\hbox{as  }\xi\to+\infty,
    \end{array}\right.
\end{veqnarray}
where
\begin{veqnarray}
  \nonumber C_+=-\frac{1}{B_1}\int d\xi\ {\rm e}^{\Lambda} G(\xi).
\end{veqnarray}

\end{theorem}

Under the conditions of Theorem~\ref{th:4.6} we also have

\begin{lemma}
  \label{th:4.7}
  
  $Q_\infty$ can be expressed as
\begin{veqnarray}
  \label{eq:4.4.5}
  Q_\infty(\xi)= -\frac{\xi{\rm e}^{-\Lambda}}{B_1}\int_{-\infty}^\xi
  \!\! d\xi' \ \xi'^{-2}{\rm e}^{\Lambda'}
  \int_{-\infty}^{\xi'}\!\!d\xi'' \ \xi'' F(\xi''),
\end{veqnarray}
for $\xi<0$, and 
\begin{veqnarray}
  \label{eq:4.4.6}
  Q_\infty(\xi)=C_+\xi{\rm e}^{-\Lambda} -\frac{\xi{\rm
      e}^{-\Lambda}}{B_1}\int_\xi^{+\infty} \!\! d\xi' \ \xi'^{-2}{\rm
    e}^{\Lambda'} \int_{\xi'}^{+\infty}\!\! d\xi'' \ \xi'' F(\xi''),
\end{veqnarray}
for $\xi>0$.

\end{lemma}

\demo{Remark 1} $\xi F(\xi)$ is integrable as a consequence of
Lemma~\ref{lem:2}. In particular (\ref{eq:4.5}) at steady state gives
\begin{veqnarray}
  \nonumber \int d\xi\ \xi F=\frac{\rho}{2}(\< s\xi_+\>+\<
  s\xi_-\>)=0.
\end{veqnarray}
\enddemo

\demo{Remark 2} $C_+$ is finite if (\ref{eq:4.4.2}) holds. To see
this, note that because of the factor ${\rm e}^{\Lambda}$, problem may
arise at $\xi=+\infty$ only.  (\ref{eq:4.4.2}) implies that we can
write
\begin{veqnarray}
  \nonumber F(\xi)= {\rm e}^{-\Lambda} f(\xi),
\end{veqnarray}
with $f(\xi)=o(\xi^2)$ as $\xi\to+\infty$. Using $\int d\xi\, \xi
F=0$, we write $G(\xi)$ as
\begin{veqnarray}
  \nonumber G(\xi)&=&F(\xi)-\frac{\xi}{B_1}
  \int_{\xi}^{+\infty}\!\!d\xi'\ \xi'
  F(\xi')\\
  &=&{\rm e}^{-\Lambda}f(\xi)-\frac{\xi}{B_1}
  \int_{\xi}^{+\infty}\!\!d\xi'\ \xi' {\rm e}^{-\Lambda'} f(\xi').
\end{veqnarray}
Using ${\rm e}^{-\Lambda'}= -B_1 \xi'^{-2} ({\rm
  e}^{-\Lambda'})_{\xi'}$, after integration by parts we have
\begin{veqnarray}
  \nonumber G(\xi)=-\xi\int_{\xi}^{+\infty}\!\!d\xi'\ 
  (\xi'^{-1}f(\xi'))_{\xi'} {\rm e}^{-\Lambda'}.
\end{veqnarray}
Since $f(\xi)=o(\xi^2)$ as $\xi\to+\infty$,
$(\xi^{-1}f(\xi))_{\xi}=o(1)$ and
\begin{veqnarray}
  \nonumber G(\xi)=\int_{\xi}^{+\infty}\!\!d\xi'\ o(1) {\rm
    e}^{-\Lambda'} =o\left(\int_{\xi}^{+\infty}\!\!d\xi'\ {\rm
      e}^{-\Lambda'}\right) =o\left(\xi^{-2}{\rm e}^{-\Lambda}\right),
\end{veqnarray}
where at the last step we used ${\rm e}^{-\Lambda}=O((\xi^{-2}{\rm
  e}^{-\Lambda})_\xi)$. Thus
\begin{veqnarray}
  \nonumber {\rm e}^{\Lambda}G(\xi)=o(\xi^{-2}) \qquad \hbox{as } \ 
  \xi\to+\infty,
\end{veqnarray}
which implies that $C_+$ is finite. Similarly, we can show that the
last integral in (\ref{eq:4.4.6}) is finite {\em iff} (\ref{eq:4.4.2})
holds. Indeed, using $F(\xi)=o(\xi^2{\rm e}^{-\Lambda})$ as
$\xi\to+\infty$ we have
\begin{veqnarray}
  \nonumber \int_{\xi'}^{+\infty}\!\! d\xi'' \ \xi'' F(\xi'')
  &=& \int_{\xi'}^{+\infty}\!\! d\xi''\ o(\xi''^3{\rm e}^{-\Lambda''})\\
  &=& o\Bigl(\int_{\xi'}^{+\infty}\!\! d\xi''\ 
  \xi''^3{\rm e}^{-\Lambda''}\Bigr)\\
  &=& o(\xi'{\rm e}^{-\Lambda'}),
\end{veqnarray}
where we used $\xi^3{\rm e}^{-\Lambda}=O((\xi{\rm
  e}^{-\Lambda})_\xi)$. Thus
\begin{veqnarray}
  \label{eq:4.4.6.b}
  \xi'^{-2}{\rm e}^{\Lambda'}\int_{\xi'}^{+\infty}\!\! d\xi'' \ \xi''
  F(\xi'')= o(\xi'^{-1}) \qquad \hbox{as } \ \xi\to+\infty,
\end{veqnarray}
which is the necessary and sufficient condition for the last integral
in (\ref{eq:4.4.6}) to be finite.
\enddemo

\begin{proof}[Proof of Lemma \ref{th:4.7}]
  
  To show (\ref{eq:4.4.5}), we start from the explicit expression for
  the integral involving $G$ in (\ref{eq:4.4.3})
\begin{veqnarray}
  \nonumber &&-\frac{\xi{\rm e}^{-\Lambda}}{B_1} \int_{-\infty}^\xi
  \!\! d\xi'
  \ {\rm e}^{\Lambda'} G(\xi')\\
  &=& -\frac{\xi{\rm e}^{-\Lambda}}{B_1} \int_{-\infty}^\xi \!\! d\xi'
  \ {\rm e}^{\Lambda'} F(\xi') -\frac{\xi{\rm e}^{-\Lambda}}{B^2_1}
  \int_{-\infty}^\xi \!\! d\xi' \ {\rm e}^{\Lambda'}
  \xi'\int_{-\infty}^{\xi'}\!\!  d\xi'' \ \xi'' F(\xi''),
\end{veqnarray}
and integrate by parts the second integral using ${\rm e}^{\Lambda'}=
B_1 \xi'^{-2} ({\rm e}^{-\Lambda'})_{\xi'}$. The result is
\begin{veqnarray}
  \nonumber &&-\frac{\xi{\rm e}^{-\Lambda}}{B_1} \int_{-\infty}^\xi
  \!\! d\xi'
  \ {\rm e}^{\Lambda'} G(\xi')\\
  &=& -\frac{1}{B_1} \int_{-\infty}^\xi \!\! d\xi' \ \xi' F(\xi')
  -\frac{\xi{\rm e}^{-\Lambda}}{B_1} \int_{-\infty}^\xi \!\! d\xi' \ 
  \xi'^{-2} {\rm e}^{\Lambda'} \int_{-\infty}^{\xi'}\!\!  d\xi'' \ 
  \xi'' F(\xi'').
\end{veqnarray}
Inserting this expression in (\ref{eq:4.4.3}) gives (\ref{eq:4.4.5}).

To show (\ref{eq:4.4.6}), we use $\int d\xi \xi F=0$ and write
(\ref{eq:4.4.3}) as
\begin{veqnarray}
  \nonumber Q_\infty(\xi)= C_+\xi{\rm e}^{-\Lambda}
  -\frac{1}{B_1}\int_{\xi}^{+\infty} \!\! d\xi' \ \xi'
  F(\xi')+\frac{\xi{\rm e}^{-\Lambda}}{B_1} \int_\xi^{+\infty} \!\!
  d\xi' \ {\rm e}^{\Lambda'} G(\xi').
\end{veqnarray} 
Using $\int d\xi \xi F=0$ we write explicitly the integral involving
$G$ in this expression as
\begin{veqnarray}
  \nonumber &&\frac{\xi{\rm e}^{-\Lambda}}{B_1} \int_\xi^{+\infty}
  \!\! d\xi'
  \ {\rm e}^{\Lambda'} G(\xi')\\
  &=& \frac{\xi{\rm e}^{-\Lambda}}{B_1} \int_\xi^{+\infty} \!\! d\xi'
  \ {\rm e}^{\Lambda'} F(\xi') -\frac{\xi{\rm e}^{-\Lambda}}{B^2_1}
  \int_\xi^{+\infty} \!\! d\xi' \ {\rm e}^{\Lambda'}
  \xi'\int_{\xi'}^{+\infty}\!\!  d\xi'' \ \xi'' F(\xi''),
\end{veqnarray}
and integrate by parts the second integral using ${\rm e}^{\Lambda'}=
B_1 \xi'^{-2} ({\rm e}^{-\Lambda'})_{\xi'}$:
\begin{veqnarray}
  \nonumber &&\frac{\xi{\rm e}^{-\Lambda}}{B_1} \int_\xi^{+\infty}
  \!\! d\xi'
  \ {\rm e}^{\Lambda'} G(\xi')\\
  &=& \frac{1}{B_1} \int_\xi^{+\infty} \!\! d\xi' \ \xi' F(\xi')
  -\frac{\xi{\rm e}^{-\Lambda}}{B_1} \int_\xi^{+\infty} \!\! d\xi' \ 
  \xi'^{-2} {\rm e}^{\Lambda'} \int_{\xi'}^{+\infty}\!\!  d\xi'' \ 
  \xi'' F(\xi'').
\end{veqnarray}
Inserting this expression in the last expression for $Q_\infty$ gives
(\ref{eq:4.4.6})

\end{proof}

\begin{proof}[Proof of Theorem \ref{th:4.6}]
  The general solution of (\ref{eq:4.4.1}) is $Q=Q_\infty +C_1 Q_1
  +C_2 Q_2$, where $C_1$ and $C_2$ are constants, $Q_1$ and $Q_2$ are
  two linearly independent solutions of the homogeneous equation
  associated with (\ref{eq:4.4.1}). Two such solutions are
\begin{veqnarray}
  \label{eq:4.4.7}
  Q_1(\xi)= \xi{\rm e}^{-\Lambda},
\end{veqnarray} 
\begin{veqnarray}
  \label{eq:4.4.8}
  Q_2(\xi)= 1- \frac{\xi{\rm e}^{-\Lambda}}{B_1} \int_{-\infty}^\xi
  \!\! d\xi' \ \xi'{\rm e}^{\Lambda'},
\end{veqnarray}
For $\xi<0$, using ${\rm e}^{\Lambda'}= B_1 \xi'^{-2} ({\rm
  e}^{\Lambda'})_{\xi'}$, after integration by parts $Q_2$ can be
written as
\begin{veqnarray}
  \label{eq:4.4.9}
  Q_2(\xi)=-\xi{\rm e}^{-\Lambda} \int_{-\infty}^\xi \!\! d\xi' \ 
  \xi'^{-2}{\rm e}^{\Lambda'}.
\end{veqnarray}

We now show that realizability requires $C_1=C_2=0$.  First, one
readily checks that $\lim_{\xi\to-\infty} Q_\infty=
\lim_{\xi\to-\infty} Q_2=0$, while $Q_1$ grows unbounded as
$\xi\to-\infty$. Hence in order that $Q$ be integrable we must set
$C_1=0$ and the general solution of (\ref{eq:4.4.1}) is
\begin{veqnarray}
  \nonumber Q(\xi)=C_2 Q_2(\xi)+Q_\infty(\xi).
\end{veqnarray}
We evaluate this solution asymptotically as $\xi\to\pm\infty$.
Consider $Q_2$ for large negative $\xi$ first. Using ${\rm
  e}^{\Lambda'}= B_1 \xi'^{-2} ({\rm e}^{\Lambda'})_{\xi'}$, after
integration by parts we write (\ref{eq:4.4.9}) as
\begin{veqnarray}
  \nonumber Q_2(\xi) =B_1|\xi|^{-3} -4B_1\xi {\rm e}^{-\Lambda}
  \int_{-\infty}^\xi \!\! d\xi' \ \xi'^{-5}{\rm e}^{\Lambda'}.
\end{veqnarray}
Since $\xi^{-5} {\rm e} ^\Lambda=O((\xi^{-7}{\rm e} ^\Lambda)_\xi)$ as
$\xi\to-\infty$, the integral in this expression is of the order
\begin{veqnarray}
  \nonumber \xi {\rm e}^{-\Lambda} \int_{-\infty}^\xi \!\! d\xi' \ 
  \xi'^{-5}{\rm e}^{\Lambda'} &=&\xi {\rm e}^{-\Lambda}
  \int_{-\infty}^\xi \!\! d\xi' O((\xi^{-7}{\rm e} ^\Lambda)_\xi)\\
  &=&O\Bigl(\xi {\rm e}^{-\Lambda}
  \int_{-\infty}^\xi \!\! d\xi' (\xi^{-7}{\rm e} ^\Lambda)_\xi\Bigr)\\
  &=&O(\xi^{-6}).
\end{veqnarray}
Thus
\begin{veqnarray}
  \nonumber Q_2(\xi)=B_1|\xi|^{-3}+O(\xi^{-6})\qquad \hbox{as } \ 
  \xi\to-\infty,
\end{veqnarray}
Consider now $Q_\infty$ for large negative $\xi$. Using ${\rm
  e}^{\Lambda'}= B_1 \xi'^{-2} ({\rm e}^{\Lambda'})_{\xi'}$, after
integration by parts we write (\ref{eq:4.4.5}) as
\begin{veqnarray}
  \nonumber
  Q_\infty(\xi)&=&|\xi|^{-3} \int_{-\infty}^\xi d\xi'\  \xi' F(\xi')\\
  &+&\xi{\rm e}^{-\Lambda} \int_{-\infty}^\xi \!\! d\xi'
  \Bigl(\xi'^{-4}\int_{-\infty}^{\xi'} d\xi''\ \xi''\ 
  F(\xi'')\Bigr)_{\xi'} {\rm e}^{\Lambda'}.
\end{veqnarray}
As $\xi\to-\infty$
\begin{veqnarray}
  \nonumber \xi{\rm e}^{-\Lambda} \int_{-\infty}^\xi \!\! d\xi'
  \Bigl(\xi'^{-4}\int_{-\infty}^{\xi'} d\xi''\ \xi''\ 
  F(\xi'')\Bigr)_{\xi'} {\rm e}^{\Lambda'} &=& \xi{\rm e}^{-\Lambda}
  \int_{-\infty}^\xi \!\! d\xi'\ o(\xi'^{-5})
  {\rm e}^{\Lambda'}\\
  &=& o\Bigl(\xi{\rm e}^{-\Lambda}\int_{-\infty}^\xi \!\! d\xi'\ 
  \xi'^{-5}{\rm e}^{\Lambda'}\Bigr)\\
  &=&o(\xi^{-6}),
\end{veqnarray}
where we used the estimate above. Thus
\begin{veqnarray}
  \nonumber Q_\infty(\xi)=|\xi|^{-3} \int_{-\infty}^\xi d\xi'\ \xi'
  F(\xi') +o(\xi^{-6})\qquad \hbox{as } \ \xi\to-\infty.
\end{veqnarray}
Combining these expressions, we have
\begin{veqnarray}
  \label{eq:4.4.10}
  Q(\xi) &=& C_2B_1|\xi|^{-3}
  + |\xi|^{-3} \int_{-\infty}^\xi d\xi'\  \xi' F(\xi')\\
  &+&O(C_2\xi^{-6})+o(\xi^{-6})\qquad \hbox{as } \ \xi\to-\infty.
\end{veqnarray}
Consider now $Q_2$ for large positive $\xi$. Write (\ref{eq:4.4.8}) as
\begin{veqnarray}
  \nonumber Q_2(\xi)= 1- \frac{\xi{\rm e}^{-\Lambda}}{B_1}
  \int_{-\infty}^{\xi_\star} \!\! d\xi' \ \xi'{\rm e}^{\Lambda'}-
  \frac{\xi{\rm e}^{-\Lambda}}{B_1} \int_{\xi_\star}^{\xi} \!\! d\xi'
  \ \xi'{\rm e}^{\Lambda'},
\end{veqnarray}
where $\xi_\star >0$ is arbitrary but fixed. Using ${\rm
  e}^{\Lambda'}= B_1 \xi'^{-2} ({\rm e}^{\Lambda'})_{\xi'}$, after
integration by parts of the second integral we get
\begin{veqnarray}
  \nonumber Q_2(\xi)&=& \xi\xi_\star^{-1}{\rm
    e}^{-\Lambda+\Lambda_\star} - \frac{\xi{\rm e}^{-\Lambda}}{B_1}
  \int_{-\infty}^{\xi_\star} \!\! d\xi' \ \xi'{\rm e}^{\Lambda'} -
  \xi{\rm e}^{-\Lambda} \int_{\xi_\star}^{\xi} \!\! d\xi'
  \ \xi'^{-2}{\rm e}^{\Lambda'}\\
  &=&- \xi{\rm e}^{-\Lambda} \int_{\xi_\star}^{\xi} \!\! d\xi' \ 
  \xi'^{-2}{\rm e}^{\Lambda'}+O(\xi{\rm e}^{-\Lambda}).
\end{veqnarray}
Integrating by parts again gives
\begin{veqnarray}
  \nonumber Q_2(\xi)=-B_1\xi^{-3} -4B_1\xi {\rm e}^{-\Lambda}
  \int_{\xi_\star}^\xi \!\! d\xi' \ \xi'^{-5}{\rm
    e}^{\Lambda'}+O(\xi{\rm e}^{-\Lambda}).
\end{veqnarray}
Since $\xi^{-5} {\rm e} ^\Lambda=O((\xi^{-7}{\rm e} ^\Lambda)_\xi)$ as
$\xi\to+\infty$, the remaining integral can be estimated as above,
yielding
\begin{veqnarray}
  \nonumber Q_2(\xi)=-B_1\xi^{-3} +O(\xi^{-6}).
\end{veqnarray}
Considering $Q_\infty$ for large positive $\xi$, we must distinguish
two cases.  If (\ref{eq:4.4.2}) does not hold, then in
(\ref{eq:4.4.3}) we decompose
\begin{veqnarray}
  \nonumber \frac{\xi{\rm e}^{-\Lambda}}{B_1} \int_{-\infty}^{\xi}
  \!\! d\xi' \ {\rm e}^{\Lambda'} G(\xi')&=& \frac{\xi{\rm
      e}^{-\Lambda}}{B_1} \int_{-\infty}^{\xi_\star} \!\! d\xi' \ {\rm
    e}^{\Lambda'} G(\xi')+\frac{\xi{\rm e}^{-\Lambda}}{B_1}
  \int_{\xi_\star}^\xi \!\! d\xi'
  \ {\rm e}^{\Lambda'} G(\xi')\\
  &=&\frac{\xi{\rm e}^{-\Lambda}}{B_1} \int_{\xi_\star}^\xi \!\! d\xi'
  \ {\rm e}^{\Lambda'} G(\xi')+O(\xi{\rm e}^{-\Lambda}).
\end{veqnarray}
and write
\begin{veqnarray}
  \nonumber Q_\infty(\xi)= \frac{1}{B_1}\int_{-\infty}^\xi \!\! d\xi'
  \ \xi' F(\xi')-\frac{\xi{\rm e}^{-\Lambda}}{B_1}
  \int_{\xi_\star}^\xi \!\! d\xi' \ {\rm e}^{\Lambda'}
  G(\xi')+O(\xi{\rm e}^{-\Lambda}).
\end{veqnarray}
The second integral in this expression is explicitly
\begin{veqnarray}
  \nonumber &&-\frac{\xi{\rm e}^{-\Lambda}}{B_1} \int_{\xi_\star}^\xi
  \!\! d\xi'
  \ {\rm e}^{\Lambda'} G(\xi')\\
  &=& -\frac{\xi{\rm e}^{-\Lambda}}{B_1} \int_{\xi_\star}^\xi \!\!
  d\xi' \ {\rm e}^{\Lambda'} F(\xi') -\frac{\xi{\rm
      e}^{-\Lambda}}{B^2_1} \int_{\xi_\star}^\xi \!\! d\xi' \ {\rm
    e}^{\Lambda'} \xi'\int_{-\infty}^{\xi'}\!\!  d\xi'' \ \xi''
  F(\xi'').
\end{veqnarray}
The integration by parts of the second integral using ${\rm
  e}^{\Lambda'}= B_1 \xi'^{-2} ({\rm e}^{-\Lambda'})_{\xi'}$ gives
\begin{veqnarray}
  \nonumber &&-\frac{\xi{\rm e}^{-\Lambda}}{B_1} \int_{\xi_\star}^\xi
  \!\! d\xi'
  \ {\rm e}^{\Lambda'} G(\xi')\\
  &=& -\frac{1}{B_1} \int_{-\infty}^\xi \!\! d\xi'
  \ \xi' F(\xi') \\
  &&-\frac{\xi{\rm e}^{-\Lambda}}{B_1} \int_{\xi_\star}^\xi \!\! d\xi'
  \ \xi'^{-2} {\rm e}^{\Lambda'} \int_{-\infty}^{\xi'}\!\!  d\xi'' \ 
  \xi'' F(\xi'')+O(\xi{\rm e}^{-\Lambda}).
\end{veqnarray}
Thus
\begin{veqnarray}
  \nonumber Q_\infty(\xi)=-\frac{\xi{\rm e}^{-\Lambda}}{B_1}
  \int_{\xi_\star}^\xi \!\! d\xi' \ \xi'^{-2} {\rm e}^{\Lambda'}
  \int_{-\infty}^{\xi'}\!\!  d\xi'' \ \xi'' F(\xi'')+O(\xi{\rm
    e}^{-\Lambda}).
\end{veqnarray}
Integrating by parts again using ${\rm e}^{\Lambda'}= B_1 \xi'^{-2}
({\rm e}^{\Lambda'})_{\xi'}$ gives
\begin{veqnarray}
  \nonumber
  Q_\infty(\xi)&=&-\xi^{-3} \int_{-\infty}^\xi\!\! d\xi'\  \xi' F(\xi')\\
  &+&\xi{\rm e}^{-\Lambda} \int_{\xi_\star}^\xi \!\! d\xi'
  \Bigl(\xi'^{-4}\int_{-\infty}^{\xi'}\!\!  d\xi''\ \xi''\ 
  F(\xi'')\Bigr)_{\xi'} {\rm e}^{\Lambda'}+O(\xi{\rm e}^{-\Lambda}).
\end{veqnarray}
The last integral can be estimate as for the large negative $\xi$.
Using $\int d\xi\,\xi F=0$, this leads to
\begin{veqnarray}
  \nonumber Q_\infty(\xi)=\xi^{-3} \int_\xi^{+\infty}\!\! d\xi'\ \xi'
  F(\xi') +o(\xi^{-6}),
\end{veqnarray}
and, hence,
\begin{veqnarray}
  \label{eq:4.4.11}
  Q(\xi) &=& -C_2B_1\xi^{-3}
  +\xi^{-3} \int_\xi^{+\infty}\!\! d\xi'\  \xi' F(\xi')\\
  &&+O(C_2\xi^{-6})+ o(\xi^{-6})\qquad \hbox{as } \ \xi\to+\infty.
\end{veqnarray}
In contrast, if (\ref{eq:4.4.2}) holds, then we can use
(\ref{eq:4.4.6}). From (\ref{eq:4.4.6.b}) it follows that the integral
term in this expression is $o(\xi{\rm e}^{-\Lambda})$ as
$\xi\to+\infty$.  Thus,
\begin{veqnarray}
  \nonumber Q(\xi)= C_+\xi{\rm e}^{-\Lambda} +o(\xi{\rm
    e}^{-\Lambda}),
\end{veqnarray}
and
\begin{veqnarray}
  \label{eq:4.4.12}
  Q_\infty(\xi)&=& -C_2B_1\xi^{-3}
  + C_+\xi{\rm e}^{-\Lambda}\\
  &&+O(C_2\xi^{-6})+ o(\xi{\rm e}^{-\Lambda})\qquad \hbox{as } \ 
  \xi\to+\infty.
\end{veqnarray}
In (\ref{eq:4.4.10}), (\ref{eq:4.4.11}), (\ref{eq:4.4.12}), if the
$C_2$ term at the right-hand side is non-zero, then it will dominate
the second term. However, since the $C_2$ term has opposite sign when
$\xi\to\pm\infty$, it must be zero, i.e., we must set $C_2=0$.  This
proves that $Q=Q_\infty$. Furthermore, since the $F$ term at the
right-hand side of (\ref{eq:4.4.11}) is negative (recall that
$F\le0$), this solution must be rejected in order that $Q$ be
non-negative.  Thus (\ref{eq:4.4.2}) must hold, and we get
(\ref{eq:4.4.4}).

\end{proof}

\demo{Remark} If the assumptions of Lemma \ref{lem:2} do not apply,
i.e. if shocks are not created at zero amplitude or shock strength
does not add up at collision, then from (\ref{eq:4.7}) at steady state
we have
\begin{veqnarray}
  \nonumber \int d\xi \ \xi F(\xi)= -D_c.
\end{veqnarray}
This implies that (\ref{eq:4.4.11}) has to be changed to
\begin{veqnarray}
  \nonumber Q(\xi) &=& -C_2B_1\xi^{-3}
  +D_c\xi^{-3}+\xi^{-3} \int_\xi^{+\infty}\!\! d\xi'\  \xi' F(\xi')\\
  &&+O(C_2\xi^{-6})+ o(\xi^{-6})\qquad \hbox{as } \ \xi\to+\infty.
\end{veqnarray}
This solution cannot be dismissed as being unrealizable since the
second term at the right hand-side can balance the first one. In
particular, for $C_2=D_c/B_1$ we obtain from (\ref{eq:4.4.10})
$Q(\xi)=D_c |\xi|^{-3}+o(\xi^{-3})$ as $\xi\to-\infty$, consistent
with (\ref{eq:4.8.1}).  \enddemo

\subsection{Statistics for the environment of the shocks}
\label{sec:4.5.1}

We now turn to the statistics for the environment of the shocks. For
simplicity we will focus on statistically homogeneous situations such
that $u(x,t)\law-u(-x,t)$. Define
\begin{veqnarray}
  \nonumber s(x,y_0,t)=u(y_0+x/2,t)-u(y_0-x/2,t),
\end{veqnarray}
and let $W(s,\xi_+,\xi_-,x,t)$ be the PDF of
\begin{veqnarray}
  \nonumber (s(x,y_0,t),\xi(y_0+x/2,t),\xi(y_0-x/2,t)),
\end{veqnarray}
conditional on $y_0$ being a shock location. Since $u(x,t)\law
-u(-x,t)$, it follows that
\begin{veqnarray}
  \nonumber W(s,\xi_+,\xi_-,x,t)= W(s,\xi_-,\xi_+,x,t).
\end{veqnarray}
$V(s,\xi,t)$ and $W(s,\xi_+,\xi_-,x,t)$ are related by (recall that
$V=V_+=V_-$ if $u(x,t)\law -u(-x,t)$)
\begin{veqnarray}
  \label{eq:4.9.l1}
  V(s,\xi,t)=\lim_{x\to0+} \int d\xi'\ W(s,\xi',\xi,x,t).
\end{veqnarray} 
Thus,
\begin{veqnarray}
   \label{eq:4.9.l2}
   F(\xi,t)=\rho\lim_{x\to0+}\int ds d\xi'\ W(s,\xi',\xi,x,t).
\end{veqnarray} 
We have

\begin{theorem}
\label{th:3.8} 
$W$ satisfies
\begin{veqnarray}
  \label{eq:4.18}
  (\rho W)_t&=& - \rho sW_x
  +2 \rho(B_0-B(x)) W_{ss}\\
  &&+\frac{\rho}{2}\xi_+W +\rho\bigl(\xi_+^2 W\bigr)_{\xi_+}
  +\frac{\rho}{2}\xi_-W
  +\rho\bigl(\xi_-^2 W\bigr)_{\xi_-}\\
  && + \rho B_1\bigl( W_{\xi_+ \xi_+} +W_{\xi_- \xi_-}\bigr)
  +2\rho  B_1(x)W_{\xi_+ \xi_-}\\
  &&- \rho B_2(x)\bigl( W_{s\xi_+}+W_{s \xi_-}\bigr)
  +\varsigma_1-\varsigma_2 +J,
\end{veqnarray}
with $B_1(x)=-B_{xx}(x)$, $B_2(x)=B_x(x)$.
$\varsigma_1(s,\xi_+,\xi_-,x,t)$ is defined such that
\begin{veqnarray}
  \nonumber \varsigma_1(s,\xi_+,\xi_-,x,t) ds d\xi_-d\xi_+ dz dt,
\end{veqnarray}
gives the average number of shock creation points in
$[z,z+dz)\times[t,t+dt)$ with
\begin{veqnarray}
  \nonumber
  s(x,y_1,t_1)&\in&[s,s+ds),\\
  \xi(y_1+x/2,t_1)&\in&[\xi_+,\xi_++d\xi_+),\\
  \xi(y_1-x/2,t_1)&\in&[\xi_-,\xi_-+d\xi_-),
\end{veqnarray}
conditional on $(y_1,t_1)\in ([z,z+dz)\times[t,t+dt))$ being a point
of shock creation (because of the statistical homogeneity, $z$ is a
dummy variable). $\varsigma_2(s,\xi_+,\xi_-,x,t)$
is defined such that
\begin{veqnarray}
  \nonumber \varsigma_2(s,\xi_+,\xi_-,x,t) ds d\xi_-d\xi_+ dz dt,
\end{veqnarray}
gives the average number of shock collision points in
$[z,z+dz)\times[t,t+dt)$ with
\begin{veqnarray}
  \nonumber
  s(x,y_2,t_2)&\in&[s,s+ds),\\
  \xi(y_2+x/2,t_2)&\in&[\xi_+,\xi_++d\xi_+),\\
  \xi(y_2-x/2,t_2)&\in&[\xi_-,\xi_-+d\xi_-),
\end{veqnarray}
conditional on $(y_2,t_2)\in ([z,z+dz)\times[t,t+dt))$ being a point
of shock collision.  Finally, $J(s,\xi_+,\xi_-,x,t)$ accounts for the
possibility of having another shock in between $[y_j-x/2,y_j+x/2]$ and
satisfies
\begin{veqnarray}
  \nonumber J(s,\xi_+,\xi_-,x,t)=O(x).
\end{veqnarray}
\end{theorem}

At statistical steady state, the definitions for $\varsigma_1$,
$\varsigma_2$ simplify. Indeed, in the limit as $t\to+\infty$, we have
\begin{veqnarray}
  \nonumber \varsigma_1(s,\xi_+,\xi_-,x,t)\to \sigma_1
  S_1(s,\xi_+,\xi_-,x),
\end{veqnarray}
where $\sigma_1$ is the space-time number density of shock creation
points, $S_1(s,\xi_+,\xi_-,x)$ is the PDF of
\begin{veqnarray}
  \nonumber
  (s(x,y_1,t_1),\xi(y_1+x/2,t_1),\xi(y_1-x/2,t_1)),
\end{veqnarray}
conditional on a shock being created at $(y_1,t_1)$, and
\begin{veqnarray}
  \nonumber \varsigma^i_2(s,\xi_+,\xi_-,x,t)\to \sigma_2
  S_2(s,\xi_+,\xi_-,x),
\end{veqnarray}
where $\sigma_2$ is the space-time number density of shock collision
points, $S_2(s,\xi_+,\xi_-,x)$ is the PDF of
\begin{veqnarray}
  \nonumber
  (s(x,y_2,t_2),\xi(y_2+x/2,t_2),\xi(y_2-x/2,t_2)),
\end{veqnarray}
conditional on a two shocks colliding at $(y_2,t_2)$.

~

\demo{Remark on the strategy for the proof of Theorem \ref{th:3.8}}
Ideally to prove Theorem \ref{th:3.8} we should follow the strategy in
Sec.~\ref{sec:4} for the derivation of the equation for $Q$.  Let
$X(u_1,x_1,\ldots,u_6,x_6,t)$ be the PDF of
\begin{veqnarray}
  \nonumber (u(y_0+x_1,t),\ldots,u(y_0+x_6,t),
\end{veqnarray}
conditional on $y_0$ being a shock location. Knowing the equation for
$X$, one can easily derive an equation for the conditional PDF of
\begin{veqnarray}
  \nonumber
  (u(y_0+x/2,t),u(y_0+x/2,t),\eta(y_0+x/2,y,t),\eta(y_0-x/2,y,t)
\end{veqnarray}
where $\eta(x,z,t)=(u(x+z,t)-u(z,t))/z$. Letting $z\to0$, one derives
an equation for $W$. Clearly, this derivation is rather tedious and,
as we now show, unnecessary for our purpose.

Recall that
\begin{veqnarray}
  \nonumber \rho X(u_1,x_1,\ldots,u_6,x_6,t)&=& \int \frac{d\lambda_1\cdots
    d\lambda_6}{(2\pi)^6}
  {\rm e}^{i\lambda_1 u_1+\cdots+i\lambda_6 u_6}\\
  &&\times \biggl\<\sum_j{\rm e}^{-i\lambda_1
    u(z+x_1,t)-\cdots-i\lambda_6 u(z+x_1,t)} \delta(z-y_j)\biggr\>.
\end{veqnarray}
The average under the integral is the characteristic function
associated with $X$, and an equation for this quantity can be derived
using the equation for $u(z+x_p,t)$
\begin{veqnarray}
  \nonumber du= -\bl u\br_\A {u}_{x_p}dt +dW(z+x_p,t), \quad
  p=1,\ldots,6.
\end{veqnarray}
Instead of reproducing these straightforward calculations, we note
simply that for homogeneous situations the resulting equation for $X$
will contain terms proportional to
\begin{veqnarray}
  \nonumber \rho_2(x_p,t)=\Bigl\<\sum_{j,k}
  \delta(x_p+z-y_k)\delta(z-y_j)\Bigr>.
\end{veqnarray}
These terms account for the probability of having another shock, say
$y_1$, between $y_0$ and $y_0+x_p$: they are the origin of $J$ in
(\ref{eq:4.18}).  Note also that technically, the $\rho_2(x_p,t)$'s
arise because of the average $\bl u\br_\A$ in the equation for
$u(z+x_p,t)$. Now, the key point is to note that
\begin{veqnarray}
  \nonumber \rho(x_p,t)=O(x_p).
\end{veqnarray}
As a direct result,
\begin{veqnarray}
  \nonumber J(s,\xi_+,\xi_-,x,t)=O(x).
\end{veqnarray}

We are eventually interested in the limit as $x\to0$ of $W$. As argued
in Sec.~\ref{sec:4.5}, in this limit the $O(x)$ terms in
(\ref{eq:4.18}) are negligible. Thus, we will not dwell on obtaining
an explicit expression for $J$. Instead, we will derive
(\ref{eq:4.18}) using ($\bar W(x,t)=W_x(x,t)$)
\begin{veqnarray}
  \label{eq:3.8.1}
  du&=& -u{u}_{x_\pm}dt +dW(z+x_\pm,t),\\
  d\xi &=& -(u{\xi}_{x_\pm}+\xi^2)dt +d\bar W(z+x_\pm,t),
\end{veqnarray}
as if no shock are present between $z$ and $z+x_\pm$. The error we are
making are accounted for by the term $J$.

\enddemo

\begin{proof}[Proof of Theorem \ref{th:3.8}]
  
  Define
\begin{veqnarray}
  \nonumber
  &&\theta(\lambda_+,\lambda_-,\mu_+,\mu_-,x_+,x_-,z,t)\\
  &=&{\rm e}^{-i\lambda_+u(z+x_+,t)-i\lambda_-u(z+x_-,t)
    -i\mu_+\xi(z+x_+,t) -i\mu_-\xi(z+x_-,t)}.
\end{veqnarray}
Then
\begin{veqnarray}
  \nonumber \rho W(s,\xi_+,\xi_-,x,t) &=&\int \frac{d\lambda
    d\mu_+d\mu_-}{(2\pi)^3} {\rm e}^{-i\lambda s -i\mu_+\xi_+
    -i\mu_-\xi_-} \\
  &&\times
  \biggl\<\sum_j\theta(\lambda,-\lambda,\mu_+,\mu_-,x/2,-x/2,z,t)
  \delta(z-y_j) \biggr\>.
\end{veqnarray}

We now derive equations for $\<\Theta\>=\<\sum _j \theta
\delta(z-y_j)\>$, $W$. We use the following rules from Ito calculus
\begin{veqnarray}
  \nonumber
  d W(x,t) d W(y,t)&=&2B(x-y) dt,\\
  d\bar W(x,t) d\bar W(y,t)&=&2B_1(x-y) dt\\
  d W(x,t) d \bar W(y,t)&=&-2B(x-y) dt,
\end{veqnarray}
where $B_1(x)=-B_{xx}(x)$, $B_2(x)=B_x(x)$.  Thus, from
(\ref{eq:3.8.1}), we obtain
\begin{veqnarray}
  \nonumber d\theta &=&
  i(\lambda_+u_+{u_+}_{x_+}+\lambda_-u_-{u_-}_{x_-} )
  \theta dt\\
  &-&(\lambda_+^2 B_0 +\lambda_-^2 B_0 +2 \lambda_+\lambda_-
  B(x_+-x_-))
  \theta dt\\
  &+&i(\mu_+u_+{\xi_+}_{x_+}
  +\mu_+\xi_+^2+\mu_-u_-{\xi_-}_{x_-}+\mu_-\xi_-^2 )
  \theta dt\\
  &-&(\mu_+^2 B_1 +\mu_-^2 B_1 +2 \mu_+\mu_- B_1(x_+-x_-))
  \theta dt\\
  &-&(\lambda_-\mu_+-\lambda_+\mu_-)B_2(x_+-x_-)
  \theta dt\\
  &-&i\lambda_+\theta  dW(z+x_+,t)-i\lambda_-\theta  dW(z+x_-,t)\\
  &-&i\mu_+\theta d\bar W(z+x_+,t)-i\mu_-\theta d\bar W(z+x_-,t),
\end{veqnarray}
where $u_\pm=u(z+x_\pm,t)$, $\xi_\pm=\xi(z+x_\pm,t)$.  Similarly, using
$dy_j/dt=\bar u(y_j,t)$, we get
\begin{veqnarray}
  \nonumber d\sum_j\delta(z-y_j)
  &=&-\sum_j\bar u(y_j,t) \delta^1(z-y_j)dt\\
  &&+\sum_k \delta(z-y_k)\delta(t-t_k)dt-\sum_l
  \delta(z-y_l)\delta(t-t_l)dt,
\end{veqnarray}
where $\delta^1(z)=d\delta(z)/dz$, the $(y_k,t_k)$'s are the points of
shock creation, the $(y_l,t_l)$'s are the points of shock collisions.  Using
\begin{veqnarray}
  \nonumber \theta\delta^1(z-y_j)=
  (\theta\delta(z-y_j))_z-\theta_{x_+}\delta(z-y_j)-
  \theta_{x_-}\delta(z-y_j),
\end{veqnarray}
and noting that $\<\cdot\>_z=0$ by statistical homogeneity, it follows
that
\begin{veqnarray}
  \nonumber \<\Theta\>_t &=& i\lambda_+\<u_+{u_+}_{x_+}\Theta \>+
  i\lambda_-\< u_-{u_-}_{x_-}\Theta \>\\
  &-&(\lambda_+^2 B_0 +\lambda_-^2 B_0 +2 \lambda_+\lambda_-
  B(x_+-x_-))
  \<\Theta\>\\
  &+&i\mu_+\< (u_+{\xi_+}_{x_+} +\xi_+^2 )\Theta\>
  +i\mu_-\< (u_-{\xi_-}_{x_-}+\xi_-^2)\Theta\>\\
  &-&(\mu_+^2 B_1 +\mu_-^2 B_1 +2 \mu_+\mu_- B_1(x_+-x_-))
  \<\Theta\>\\
  &-&(\lambda_-\mu_+-\lambda_+\mu_-)B_2(x_+-x_-)
  \<\Theta\>\\
  &+&\biggl\<\sum_j\bar u(y_j,t)
  (\theta_{x_+}+\theta_{x_-})\delta(z-y_j)\biggr\>
  +{\it\Sigma}_1-{\it\Sigma}_2,
\end{veqnarray}
where ${\it\Sigma}_1$ and ${\it\Sigma}_2$ account respectively for
shock creation and collision events. These are given by
\begin{veqnarray}
  \nonumber
  &&{\it \Sigma}_1(\lambda_+,\lambda_-,\mu_+,\mu_-,x_+,x_-,t)\\
  &=&\biggl\<\sum_k {\rm e}^{-i\lambda_+ u_+-i\lambda_- u_-
    -i\mu_+\xi_+-i\mu_-\xi_-}\delta(z-y_k)\delta(t-t_k) \biggr\>,
\end{veqnarray}
\begin{veqnarray}
  \nonumber
  &&{\it \Sigma}_2(\lambda_+,\lambda_-,\mu_+,\mu_-,x_+,x_-,t)\\
  &=&\biggl\<\sum_l {\rm e}^{-i\lambda_+ u_+-i\lambda_- u_-
    -i\mu_+\xi_+-i\mu_-\xi_-}\delta(z-y_l)\delta(t-t_l) \biggr\>.
\end{veqnarray}
To average the convective terms we use:
\begin{veqnarray}
  \nonumber &&i\lambda_\pm\< u_\pm{u_\pm}_{x_\pm}\Theta\>
  +i\mu_\pm\<(u_\pm{\xi_\pm}_{x_\pm}
  +\xi_\pm^2 )\Theta\>\\
  &=&-\<u_\pm\Theta_{x_\pm}\>
  -i\mu_\pm\<\Theta\>_{\mu_\pm\mu_\pm}\\
  &=&-\<u_\pm\Theta\>_{x_\pm}+ \<\xi_\pm\Theta\>
  -i\mu_\pm\<\Theta\>_{\mu_\pm\mu_\pm}\\
  &=&-i\<\Theta\>_{x_\pm\lambda_\pm}+i\<\Theta\>_{\mu_\pm}
  -i\mu_\pm\<\Theta\>_{\mu_\pm\mu_\pm}.
\end{veqnarray}
For the term involving $\bar u(y_j,t)=(u_+(y_j,t)+u_-(y_j,t))/2$ we
note that
\begin{veqnarray}
  \nonumber u_+(y_j,t)\theta _{x_+}&=& u(y_j+x_+,t)\theta _{x_+}
  -x_+\int_0^1\!\! d\beta\ \xi(y_j+\beta x_+,t) \theta _{x_+}\\
  &=& (u(y_j+x_+,t)\theta)_{x_+}-\xi(y_j+x_+,t)\theta\\
  &&-x_+\int_0^1\!\! d\beta\ \xi(y_j+\beta x_+,t) \theta _{x_+}\\
  &=& i\theta _{x_+\lambda_+}-i\theta _{\mu_+} -x_+\int_0^1\!\!
  d\beta\ \xi(y_j+\beta x_+,t) \theta _{x_+}.
\end{veqnarray}
A similar expression holds for $u_-(y_j,t)\theta _{x_-}$. Also
\begin{veqnarray}
  \nonumber u_+(y_j,t)\theta _{x_-}&=& (u(y_j+x_+,t)\theta_j)_{x_-}
  -x_+\int_0^1\!\! d\beta\ \xi(y_j+\beta x_+,t) \theta _{x_-}\\
  &=& i\theta _{x_-\lambda_+} -x_+\int_0^1\!\! d\beta\ \xi(y_j+\beta
  x_+,t) \theta _{x_-},
\end{veqnarray}
and a similar expression holds for $u_-(y_j,t)\theta _{x_+}$. Thus
\begin{veqnarray}
  \nonumber &&2\biggl\<\sum_j\bar u(y_j,t)
  (\theta_{x_+}+\theta_{x_-})\delta(z-y_j)
  \biggr\>\\
  &=&i \<\Theta\>_{x_+\lambda_+}+i \<\Theta\>_{x_-\lambda_+}
  +i\<\Theta\>_{x_+\lambda_-}+i\<\Theta\>_{x_-\lambda_-}
  -i\<\Theta\>_{\mu_+} -i\<\Theta\>_{\mu_-}-R
\end{veqnarray}
where
\begin{veqnarray}
  \nonumber
  &&R(\lambda_+,\lambda_-,\mu_+,\mu_-,x_+,x_-)\\
  &=& x_+\biggl\<\sum_j \int_0^1\!\! d\beta\ \xi(y_j+\beta x_+,t)
  (\theta _{x_+}+\theta _{x_-})\delta(z-y_j)\biggr\>\\
  &+&x_-\biggl\<\sum_j \int_0^1\!\! d\beta\ \xi(y_j+\beta x_-,t)
  (\theta _{x_+}+\theta _{x_-})\delta(z-y_j)\biggr\>.
\end{veqnarray}
Combining the above expressions leads to the following equation
for~$\<\Theta\>$:
\begin{veqnarray}
  \nonumber \<\Theta\>_t&=& -\frac{i}{2} (\<\Theta\>_{x_+\lambda_+}
  +\<\Theta\>_{x_-\lambda_-}
  -\<\Theta\>_{x_+\lambda_-}-\<\Theta\>_{x_-\lambda_+})\\
  &&-(\lambda_+^2 B_0 +\lambda_-^2 B_0 +2 \lambda_+\lambda_-
  B(x_+-x_-))
  \<\Theta\>\\
  &&+\frac{i}{2}(\<\Theta\>_{\mu_+}+\<\Theta\>_{\mu_-})
  -i\mu_+\<\Theta\>_{\mu_+\mu_+}-i\mu_-\<\Theta\>_{\mu_-\mu_-}\\
  &&-(\mu_+^2 B_1 +\mu_-^2 B_1 +2 \mu_+\mu_- B_1(x_+-x_-))
  \<\Theta\>\\
  &&-(\lambda_-\mu_+-\lambda_+\mu_-)B_2(x_+-x_-)
  \<\Theta\> +{\it\Sigma}_1-{\it\Sigma}_2-R.
\end{veqnarray}
To obtain an equation for $W$, we note the following remarkable
property of $R$:

\begin{lemma}
\begin{veqnarray}
  \nonumber R(\lambda_+,\lambda_-,\mu_+,\mu_-,x/2,-x/2)=0.
\end{veqnarray}
\end{lemma}

\begin{proof}
  To see this, write
\begin{veqnarray}
  \nonumber
  && R(\lambda_+,\lambda_-,\mu_+,\mu_-,x/2,-x/2)\\
  &=&\frac{x}{2} \lim_{\bar x\to0}\frac{\partial}{\partial\bar x}
  \biggl\<\sum_j \int_0^1\!\! d\beta\ (\xi(y_j+\beta
  x/2,t)-\xi(y_j-\beta x/2,t))\theta\delta(z-y_j)\biggr\>,
\end{veqnarray}
where $\theta= \theta_j(\lambda_+,\lambda_-,\mu_+,\mu_-,\bar
x+x/2,\bar x-x/2,z,t)$.  We claim that
\begin{veqnarray}
  \nonumber A&=&\biggl\<\sum_j \int_0^1\!\! d\beta\ (\xi(y_j+\beta
  x/2,t)-\xi(y_j-\beta x/2,t))\theta\delta(z-y_j)\biggr\>=0.
\end{veqnarray}
Indeed the symmetry $u(x,t)\law -u(-x,t)$ requires that $A$ be
invariant under the transformation
\begin{veqnarray}
  \nonumber z\to-z,\quad x\to x, \quad \bar x\to -\bar x, \quad y_j\to
  -y_j, \quad \lambda_\pm \to -\lambda_\mp, \quad \mu_\pm \to \mu_\mp.
\end{veqnarray}
On the other hand one checks explicitly that $A\to-A$ under the same
transformation. Hence $A=0$, $R=0$.
\end{proof}

We now continue with the proof of Lemma~\ref{th:3.8}. Combining the
above expressions, on the subset $\lambda_+=\lambda$,
$\lambda_-=-\lambda$, $x_+=x/2$, $x_-=-x/2$, $\<\Theta\>$ satisfies
\begin{veqnarray}
  \nonumber \<\Theta\>_t&=& -i \<\Theta\>_{x\lambda}-2\lambda^2 (B_0 -
  B(x))
  \<\Theta\>\\
  &&+\frac{i}{2}(\<\Theta\>_{\mu_+}+\<\Theta\>_{\mu_-})
  -i\mu_+\<\Theta\>_{\mu_+\mu_+}-i\mu_-\<\Theta\>_{\mu_-\mu_-}\\
  &&-(\mu_+^2 B_1 +\mu_-^2 B_1 +2 \mu_+\mu_- B_1(x))
  \<\Theta\>\\
  &&+(\lambda\mu_++\lambda\mu_-)B_2(x)
  \<\Theta\> +{\it\Sigma}_1-{\it\Sigma}_2,
\end{veqnarray}
where the ${\it\Sigma}_1$, ${\it\Sigma}_2$ are evaluated at
$\lambda_+=\lambda$, $\lambda_-=-\lambda$, $x_+=x/2,x_-=-x/2$. Going
to the variables $(s,\xi_+,\xi_-)$ we obtain (\ref{eq:4.18}).

\end{proof}

\subsection{The exponent $7/2$}
\label{sec:4.5}

In this section we will derive the following result:

~

{\em
\noindent For  large negative $\xi$, $F$ behaves as
  \begin{veqnarray}
    \label{eq:4.16}
    F(\xi)\sim C |\xi|^{-5/2}\qquad \hbox{as } \xi\to-\infty.
  \end{veqnarray}
  }

~

A direct consequence of (\ref{eq:4.16}) is
\begin{veqnarray}
  \label{eq:4.17}
  Q_\infty(\xi)\sim \left\{
    \begin{array}{ll}
      {\d C_-|\xi|^{-7/2} 
        \qquad}&\hbox{as  } \xi\to-\infty,\\
      {\d C_+ \xi{\rm e}^{-\Lambda}} 
        &\hbox{as  }\xi\to+\infty,
    \end{array}\right.
\end{veqnarray}

The argument for (\ref{eq:4.16}) is based on the following result

~

{\em Consider 
\begin{veqnarray}
  \nonumber x^{-1} S_1(\bar s x^{1/3},\bar\xi_+
  x^{-2/3},\bar\xi_-x^{-2/3},x),
\end{veqnarray}
the PDF of the rescaled variables
\begin{veqnarray}
  \nonumber
  (s(y_1,x,t_1)x^{-1/3},\xi_+(y_1+x/2,t_1)x^{2/3},\xi_-(y_1-x/2,t_1)x^{2/3}),
\end{veqnarray}
conditional on a shock being created at $(y_1,t_1)$. In the limit as
$x\to0$,
\begin{veqnarray}
  \label{eq:4.19}
  &&x^{-1} S_1(\bar s x^{1/3},\bar\xi_+ x^{-2/3},\bar\xi_-x^{-2/3},x)\\
  &&\qquad\qquad\to P(\bar s) \delta\left(\bar\xi_+-\bar
    s/3\right)\delta(\bar\xi_+-\bar\xi_-),
\end{veqnarray}
where $P(\cdot)$ is a PDF supported on $(-\infty,0]$.  }

~

(\ref{eq:4.19}) shows that, in the original variables, $S_1$ is
asymptotically
\begin{veqnarray}
  \label{eq:4.19b}
  &&S_1(s,\xi_+,\xi_-,x)\\
  &\sim& x^{-1/3} P(s x^{1/3})
  \delta(\xi_+-sx^{-1}/3)\delta(\xi_+-\xi_-).
\end{veqnarray}

We first derive (\ref{eq:4.16}), then (\ref{eq:4.19}).

\demo{Derivation of (\ref{eq:4.16})} Recall that
\begin{veqnarray}
  \nonumber F(\xi)=\rho\lim_{x\to0+}\int ds d\xi'\ 
  W_\infty(s,\xi',\xi,x),
\end{veqnarray} 
where $W_\infty$ is the statistical steady state value of $W$.  Thus,
evaluating $F$ amounts to evaluating $W$ which we will do by analyzing
(\ref{eq:4.18}). We note first that this equation describe the process
of shock creation, motion, then collision. Since collision only occurs
if shocks are present, it is natural to represent the effect of
collision as proportional to the density of shocks in the systems,
i.e. we write
\begin{veqnarray}
  \nonumber \varsigma_2=\rho
  g(s,\xi_+,\xi_-,x)W,
\end{veqnarray}
for some function $g(s,\xi_-,\xi_+,x)$ which is assumed to be smooth
in $s$, $\xi_\pm$, $x$. This amounts to assuming that the
characteristics of the shocks around the collision points are not very
different from the characteristics of the shocks away from the
collision points. For instance, if they were identical, we would have
$\varsigma_2=\sigma_2 W$ and, hence, $g=\sigma_2/\rho$. Next, since we
are interested in $W$ evaluated at $x=0$, we neglect the terms
\begin{veqnarray}
  \nonumber B_2(x)=O(x), \qquad 2(B_0-B(x))=O(x^2)\qquad J=O(x),
\end{veqnarray}
in (\ref{eq:4.18}). Finally, since we are interested in the limit as
$\xi_\pm\to-\infty$, we neglect the forcing terms in $\xi_\pm$,
proportional to $B_1$ or $B_1(x)$.

Under these approximations, (\ref{eq:4.18}) reduces to
\begin{veqnarray}
  \nonumber (\rho W)_t=- \rho sW_x +\frac{\rho}{2}(\xi_++\xi_-)W
  +\rho\bigl(\xi_+^2 W\bigr)_{\xi_+} +\rho\bigl(\xi_-^2
  W\bigr)_{\xi_-} -\rho g W+\varsigma_1.
\end{veqnarray}
Assuming no shocks are present at the initial time, the equation must
be solved with the initial condition $\rho W(s,\xi_+,\xi_-,x,0)=0$.
The solution is
\begin{veqnarray}
  \nonumber \rho W(s,\xi_+,\xi_-,x,t)&=&\int_0^t
  d\tau  \ ((1-\xi_+\tau )(1-\xi_-\tau ))^{-5/2} \\
  &&\times\exp\left(-\int_0^\tau \!\! d\tau '\ 
    g\left(s,\frac{\xi_+}{1-\xi_+\tau '}%
      ,\frac{\xi_-}{1-\xi_-\tau '},x-\tau 's\right)
  \right)\\
  &&\times \varsigma_1\left(s,%
    \frac{\xi_+}{1-\xi_+\tau },\frac{\xi_-}{1-\xi_-\tau }, x-\tau
    s,t-\tau \right).
\end{veqnarray}
The statistical steady state solution is obtained in the limit as
$t\to\infty$ of this expression. Using
\begin{veqnarray}
  \nonumber \lim_{t\to\infty} \varsigma_1(s,\xi_+,\xi_-,x,t)=\sigma_1
  S_1(s,\xi_+,\xi_-,x),
\end{veqnarray}
we obtain
\begin{veqnarray}
  \nonumber \rho W_\infty(s,\xi_+,\xi_-,x)&=&\sigma_1\int_0^\infty
  d\tau  \ ((1-\xi_+\tau )(1-\xi_-\tau ))^{-5/2} \\
  &&\times\exp\left(-\int_0^\tau \!\! d\tau '\ 
     g\left(s,\frac{\xi_+}{1-\xi_+\tau '},\frac{\xi_-}{1-\xi_-\tau '},%
       x-\tau 's\right)
   \right)\\
   &&\times S_1\left(s,\frac{\xi_+}{1-\xi_+\tau
       },\frac{\xi_-}{1-\xi_-\tau }, x-\tau s\right).
\end{veqnarray}
At $x=0$, using (\ref{eq:4.19b}) for $S_1$, we get
\begin{veqnarray}
  \nonumber
  &&\rho W_\infty(s,\xi_+,\xi_-,0)\\
  &=&\sigma_1\int_0^\infty
  d\tau  \ ((1-\xi_+\tau )(1-\xi_-\tau ))^{-5/2} \\
  &\times&\exp\left(-\int_0^\tau \!\! d\tau '\ 
     g\left(s,\frac{\xi_+}{1-\xi_+\tau '},\frac{\xi_-}{1-\xi_-\tau '},%
       -\tau 's\right)
   \right)\\
   &\times& (|s|\tau )^{-1/3} P\left(-\frac{s^{2/3}}{\tau
       ^{1/3}}\right) \delta\left(\frac{\xi_+}{1-\xi_+\tau
       }+\frac{1}{3\tau }\right) \delta\left(\frac{\xi_+}{1-\xi_+\tau
       }- \frac{\xi_-}{1-\xi_-\tau }\right).
\end{veqnarray}
Since
\begin{veqnarray}
  \nonumber \delta\left(\frac{\xi_+}{1-\xi_+\tau }-
    \frac{\xi_-}{1-\xi_-\tau }\right) = (1-\xi_+\tau )^2
  \delta(\xi_+-\xi_-),
\end{veqnarray}
we have $W_\infty(s,\xi_+,\xi_-,0)\propto \delta(\xi_+-\xi_-)$.  Using
the relation (\ref{eq:4.9.l1}) between $W$ and $V$, this implies that
$W_\infty(s,\xi_+,\xi_-,0)=V_\infty(s,\xi_+)\delta(\xi_+-\xi_-)$, and
leads to
\begin{veqnarray}
  \nonumber \rho V_\infty(s,\xi)&=&\sigma_1\int_0^\infty d\tau \ 
  (1-\xi \tau )^{-3} (|s|\tau )^{-1/3} P\biggl(-\frac{s^{2/3}}{\tau
    ^{1/3}}\biggr)
  \delta\left(\frac{\xi}{1-\xi \tau }+\frac{1}{3\tau }\right)\\
  &&\times\exp\left(-\int_0^\tau \!\! d\tau '\ 
     g\left(s,\frac{\xi}{1-\xi \tau '},\frac{\xi }{1-\xi \tau '},%
       -\tau 's\right) \right).
\end{veqnarray}
To perform the integration over $\tau $ we use
\begin{veqnarray}
  \nonumber \delta\left(\frac{\xi}{1-\xi \tau }+\frac{1}{3\tau
      }\right)= \frac{9}{8\xi^2}\delta\left(\tau
    +\frac{1}{2\xi}\right).
\end{veqnarray}
Since we are considering $\xi\ll -1$, the exponential factor evaluated
at $\tau =-1/2\xi$ is
\begin{veqnarray}
  \nonumber \exp\left(-\int_0^{-1/2\xi}\!\! d\tau '\ 
     g\left(s,\frac{\xi}{1-\xi \tau '},\frac{\xi }{1-\xi \tau '},%
       -\tau 's\right) \right)=1 +O(\xi^{-1}).
\end{veqnarray}
This means that shock collision events make no contribution to leading
order, leaving us with
\begin{veqnarray}
  \nonumber \rho V_\infty(s,\xi)= \bar C \sigma_1 |s| ^{-1/3} |\xi|
  ^{-5/3} P\left(-(2s^2|\xi|)^{1/3}\right),
\end{veqnarray}
where $\bar C=2^{4/3}/3$. Hence,
\begin{veqnarray}
  \label{eq:4.19.b}
  F(\xi) = -\bar C \sigma_1 \int_{-\infty}^0\!\! ds \ s ^{2/3} |\xi|
  ^{-5/3} P\left(-(2s^2|\xi|)^{1/3}\right) = -C |\xi|^{-5/2},
\end{veqnarray}
where $C = 2^{-1/2}\sigma_1 \int_{-\infty}^0 db \, |b|^{3/2} P(b)$.
\enddemo

~

\demo{Derivation of (\ref{eq:4.19})} We use local analysis around the
shock creation points \cite{fofr83}.  Consider a shock created at
$y_1$ at time $t_1$ with velocity $u_1=u(y_1,t_1)$.  Assuming $x$ is
an analytical function of $u$, we have locally
\begin{veqnarray}
  \label{eq:3.31.0}
  x=a(u(y_1+x/2,t_1)-u_1)^3+O(u(y_1+x/2,t_1)-u_1)^4),
\end{veqnarray}
where $a\le0$ is a random quantity.  Setting $a=(2/b)^3$ gives
\begin{veqnarray}
  \nonumber u(y_1+x/2,t_1)&=&u_1+\frac{b}{2} x^{1/3} +O(x^{2/3}).
\end{veqnarray}
Hence
\begin{veqnarray}
  \label{eq:4.9.1}
  s(y_1,x,t_1)&=&u(y_1+x/2,t_1)-u(y_1-x/2,t_1)\\
  &=&b x^{1/3}+O(x^{2/3}),
\end{veqnarray}
and
\begin{veqnarray}
  \label{eq:4.9.2}
  \xi(y_1+x/2,t_1)=\frac{b}{3} x^{-2/3}+O(x^{-1/3}).
\end{veqnarray} 
Note that these formulae are only valid if there is no other shock in
$[y_1-x/2,y_1+x/2]$. Since the probability of having another shock in
$[y_1-x/2,y_1+x/2]$ is at most $O(x)$, the errors we incur by using
(\ref{eq:4.9.1})-(\ref{eq:4.9.2}) are of higher order.

Recall that
\begin{veqnarray}
  \nonumber S_1(s,\xi_+,\xi_-,x)= \int\frac{d\lambda
    d\mu_+d\mu_-}{(2\pi)^3} {\rm e}^{i\lambda
    s+i\mu_+\xi_++i\mu_-\xi_-}\Omega(\lambda,\mu_+,\mu_-,x),
\end{veqnarray}
where
\begin{veqnarray}
  \nonumber
  &&\sigma_1\Omega(\lambda,\mu_+,\mu_-,x) \\
  &=&\biggl\<\sum_k {\rm e}^{-i\lambda
    s(y_k,x,t_k)-i\mu_+\xi(y_k+x/2,t_k)
    -i\mu_-\xi(y_k-x/2,t_k)}\delta(z-y_k)\delta(t-t_k) \biggr\>.
\end{veqnarray}
$\Omega$ is the characteristic function associated with $S_1$.
Similarly
\begin{veqnarray}
  \nonumber \Omega(\bar\lambda
  x^{-1/3},\bar\mu_+x^{2/3},\bar\mu_-x^{2/3},x),
\end{veqnarray}
is the characteristic function associated with the rescaled PDF
\begin{veqnarray}
  \nonumber x S_1(\bar
  sx^{1/3},\bar\xi_+x^{-2/3},\bar\xi_-x^{-2/3},x).
\end{veqnarray}
We evaluate $\Omega(\bar\lambda
x^{-1/3},\bar\mu_+x^{2/3},\bar\mu_-x^{2/3},x)$ in the limit as $x\to0$
using (\ref{eq:4.9.1}) and (\ref{eq:4.9.2}) for $s(y_1,x,t_1)$,
$\xi(y_1+x/2,t_1)$.  This gives
\begin{veqnarray}
  \nonumber &&\sigma_1\Omega(\bar\lambda
  x^{-1/3},\bar\mu_+x^{2/3},\bar\mu_-x^{2/3},x)\\
  &=& \biggl\<\sum_k{\rm e}^{-i\bar\lambda b -
    i(\bar\mu_++\bar\mu_-)b/3}
  \delta(z-y_k)\delta(t-t_k)\biggr\>+O(x^{1/3}).
\end{veqnarray}
In the limit as $x\to0$, $b$, $(y_k,t_k)$ are the only random
quantities to be averaged over. Furthermore $b$ is statistically
independent of $(y_k,t_k)$ because of statistical homogeneity and
stationarity. Let $P(b)$ be the PDF of $b$. Then
\begin{veqnarray}
  \nonumber &&\sigma_1\lim_{x\to0}\Omega(\bar\lambda
  x^{-1/3},\bar\mu_+x^{2/3},\bar\mu_-x^{2/3},x)\\
  &=& \biggl\<\sum_k \delta(z-y_k)\delta(t-t_k)\biggr\>
  \int_{-\infty}^0\!\! db \ P(b){\rm e}^{-i\bar\lambda
    b -i(\bar\mu_++\bar\mu_-)b/3}\\
  &=& \sigma_1 \int_{-\infty}^0\!\! db \ P(b){\rm e}^{-i\bar\lambda b
    -i(\bar\mu_++\bar\mu_-)b/3}.
\end{veqnarray} 
Direct evaluation of this expression gives (\ref{eq:4.19}) in the
variables $(\bar s,\bar\xi_+,\bar\xi_-)$.

\enddemo

\subsection{Connection with the geometric picture}
\label{sec:4.6}

Here we compute directly the contribution to $F$ in the neighborhood
of shock creation. This is a reformulation of the argument presented
in \cite{ekhma97} in terms of quantities defined in the present paper.
Assume a shock is created at time $t=0$, position $x=y_1$, and with
velocity $u=u_1$. Then locally (compare (\ref{eq:3.31.0}))
\begin{veqnarray}
  \nonumber x=y_1+(u-u_1)t+a(u-u_1)^3+O((u-u_1)^2t),
\end{veqnarray}
where $a\le0$ is a random quantity. For the purpose of comparison with
(\ref{eq:4.19.b}), it is useful to set $a=(2/b)^3$.  Since for $t\ll1$
to leading order the shock is located at $x=y_1$, to leading order
$u_-(y_1,t)$, $u_+(y_1,t)$ are solution of
$0=(u-u_1)t-(2(u-u_1)/b)^3$.  Thus
\begin{veqnarray}
  \nonumber &&u_\pm(y_1,t)=u_1\mp
  \Bigl(\frac{|b|^3t}{8}\Bigr)^{1/2}+O(t),
\end{veqnarray}
\begin{veqnarray}
  \label{eq:4.31}
  &&s(y_1,t) = -\Bigl(\frac{|b|^3t}{2}\Bigr)^{1/2}+O(t)
\end{veqnarray}
Similarly, to leading order $\xi_-(y_1,t)$, $\xi_+(y_1,t)$ are
solution of $1=\xi t-3(2/b)^3(u_\pm-u_1)^2\xi$. Thus
\begin{veqnarray}
    \label{eq:4.32}
    \xi_\pm (y_1,t) = -\frac{1}{2t}+O(1).
\end{veqnarray}
Recall that from (\ref{eq:4.9.l2}) (using $\xi_+(x,t)\law
\xi_-(-x,t)$)
\begin{veqnarray}
  \nonumber F(\xi)&=&\int\frac{d\mu}{2\pi} {\rm
    e}^{i\mu\xi}\biggl\<\sum_j s(z,t) {\rm
    e}^{-i\mu\xi_+(z,t)}\delta(z-y_j) \biggr\>\\
  &=&\biggl\<\sum_j s(z,t) \xi-\xi_+(z,t)\delta(z-y_j) \biggr\>,
\end{veqnarray}
Under the assumption of ergodicity with respect to
time-translation,  $F(\xi)$ can be  evaluated  from
\begin{veqnarray}
  \nonumber F(\xi)&=&\lim_{L,T\to+\infty}\frac{1}{2LT} \int_0^T\!\!dt
  \int_{-L}^{L}dz
  \sum_{j} s(z,t) \xi-\xi_+(z,t)\delta(z-y_j)\\
  &=&\lim_{L,T\to+\infty}\frac{1}{2LT} \int_0^T\!\!dt
  \sum_{j=1}^{N(t)} s(y_j,t) \xi-\xi_+(y_j,t),
\end{veqnarray}
where $N(t)$ is number of shocks in $[-L,L]$ at time $t$.  The
contribution to $F$ near shock creation points, say $F_1$, can be
evaluated for large negative $\xi$ using (\ref{eq:4.31}),
(\ref{eq:4.32}) for $s(y_1,t)$, $\xi_+(y_1,t)$. This gives in the
limit as $\xi\to-\infty$
\begin{veqnarray}
  \nonumber F_1(\xi)\sim -\sigma_1\int_{-\infty}^0\!\!db
  P(b)\int_0^{+\infty} \!\! dt \ 
  \Bigl(\frac{|b|^3t}{2}\Bigr)^{1/2} \delta
  \left(\xi+\frac{1}{2t}\right)=-C |\xi|^{-5/2},
\end{veqnarray}
where $C = 2^{-1/2} \sigma_1\int_{-\infty}^0 db \, |b|^{3/2} P(b)$.
Comparing with (\ref{eq:4.19.b}), we conclude that $F_1(\xi)= F(\xi)$
to leading order.

\section{Conclusions}
\label{sec:5}

To recapitulate the highlights of this paper, by writing down and
working with the master equations in the inviscid limit, we have shown
that the scaling of the structure functions is related to the shocks
which are the singular structure in the limiting flow. The scaling of
the PDFs, on the other hand, is related to the shock creation and
collision points, which are singularities on the singular structures.

The present paper provides a framework within which various
statistical quantities of the stochastic Burgers equation can be
calculated using self-consistent asymptotics without making closure
assumptions.  The main examples used here are the asymptotic behavior
of structure functions and the PDF of the velocity gradient. It seems
likely that other statistical quantities, such as the tails of the
velocity PDF and the PDF for velocity difference, can also be analyzed
in the present framework by exploiting further the source terms in
(\ref{eq:3.6}), (\ref{eq:3.20}).

\appendix           

\appdx[Master equations for the viscous case]
\label{sec:appendix}

In this Appendix we list results for the PDFs in the viscous case. The
master equations below were previously derived, e.g., in
\cite{gokr93,gokr98,pol95}.

Let $P^\nu(u,\xi,x,t)$ be the PDF of $(u(x,t),\xi(x,t))$ for solutions
of (\ref{eq:1.1}). First we have

\begin{lemma}
\label{th:A.1}
$P^\nu$ satisfies
\begin{veqnarray}
  \label{eq:A.1}
  P^\nu_t&=&-u P^\nu_x+\xi P+\bigl(\xi^2 P^\nu\bigr)_\xi
  +B_0 P^\nu_{uu}+B_1 P^\nu_{\xi\xi}\\
  &&-\nu\bigl( \< u_{xx}|u,\xi\> P^\nu\bigr)_u -\nu\bigl( \<
  \xi_{xx}|u,\xi\> P^\nu\bigr)_\xi,
\end{veqnarray}
where $B_0=B(0)$, $B_1=-B_{xx}(0)$, $\<\cdot|u,\xi\>$ denotes the
conditional average on $u$ and $\xi$.
\end{lemma}

(\ref{eq:A.1}) is unclosed since the form of $\< u_{xx}|u,\xi\>$ and
$\< \xi_{xx}|u,\xi\>$ entering the viscous terms is unknown. Most work
has resorted to various closure assumptions. Our main goal has been to
find ways to extract information from the master equations such as
(\ref{eq:A.1}), without making any closure assumption. Note that from
the identity
\begin{veqnarray}
  \nonumber P^\nu_{xx}&=&-(\< u_{xx}|u,\xi\> P^\nu)_u
  -(\< \xi_{xx}|u,\xi\> P^\nu)_\xi\\
  &&+(\< u^2_x|u,\xi\> P^\nu)_{uu} +2(\< u_x\xi_x|u,\xi\>
  P^\nu)_{u\xi} +(\< \xi^2_x|u,\xi\> P^\nu)_{\xi\xi},
\end{veqnarray}
the viscous term in (\ref{eq:A.1}) can also be written as (using $\<
u^2_x|u,\xi\>=\xi^2$, $\< u_x\xi_x|u,\xi\>=\xi\<\xi_x|u,\xi\>$)
\begin{veqnarray}
  \nonumber &&-\nu\bigl(\< u_{xx}|u,\xi)\> P^\nu\big)_u
  -\nu\bigl(\< \xi_{xx}|u,\xi)\> P^\nu\big)_\xi\\
  &=& \nu P^\nu_{xx} -\nu\xi^2 P^\nu_{uu} -\nu\bigl(\< \xi_x^2|u,\xi\>
  P^\nu\bigr)_{\xi\xi} -2\nu\bigl(\xi \< \xi_x |u,\xi\>
  P^\nu\bigr)_{u\xi}.
\end{veqnarray}
Thus, viscous effects give rise to {\em anti-diffusion} terms since
$\xi^2\geq0$ and $\< \xi_x^2|u,\xi\>\geq0$: this is natural since
viscosity tends to shrink the distribution $P^\nu$ towards the origin.

Since $\xi=u_x$, we have $\< a(u)\>_x= \< a_u(u)\xi\>$ for all smooth
and compactly supported functions $a(\cdot)$. This is expressed as

\begin{lemma}
\label{th:A.2}
The consistency relation
\begin{veqnarray}
  \label{eq:A.3}
  \int d\xi\ P^\nu_x+\int d\xi\ \xi P^\nu_u=0,
\end{veqnarray}
holds for all time for the solution of (\ref{eq:A.1}) if it holds
initially.
\end{lemma}

Lemma~\ref{th:A.2} can be proven upon noting that $A=\int d\xi\,
P^\nu_x+\int d\xi\, \xi P^\nu_u$ satisfies $A_t=-u A_x +B_0 A_{uu}$,
an equation that can obtained by integration of (\ref{eq:A.1}).  Since
$A\equiv0$ initially, it is zero for all time. In the statistically
homogeneous case, (\ref{eq:A.3}) reduces to $\int d\xi\, \xi
P^\nu_u=0$ (or equivalently $\< a(u)\>_x= \< a_u(u)\xi\>=0$).
(\ref{eq:A.3}) also ensures that this equation preserves the
normalization of $P^\nu$. In fact we have

\begin{corollary}
\label{th:A.3}
The solution of (\ref{eq:A.1}) satisfies
\begin{veqnarray}
  \label{eq:A.4}
  \frac{d}{dt} \int du d\xi\ P^\nu = 0.
\end{veqnarray}
i.e. $\int du d\xi\, P^\nu = \int du d\xi\, P_0$.
\end{corollary}

(\ref{eq:A.4}) follows immediately from integrating (\ref{eq:A.1}):
\begin{veqnarray}
  \nonumber \frac{d}{dt} \int du d\xi\ P^\nu = -\< u_x\>+\<\xi\>
  +\hbox{boundary terms}=0.
\end{veqnarray}
The first two terms at the right hand-side cancel because of
(\ref{eq:A.4}); the boundary terms vanish because for finite $\nu$,
$P^\nu$ decays faster than algebraically in $\xi$ as
$|\xi|\to+\infty$.

Consider the reduced distributions
\begin{veqnarray}
  \nonumber Q^\nu(\xi,x,t) = \int du\ P^\nu(u,\xi,x,t),
\end{veqnarray} 
\begin{veqnarray}
  \nonumber R^\nu(u,x,t) = \int d\xi\ P^\nu(u,\xi,x,t).
\end{veqnarray}
(\ref{eq:A.3}), written after integration over $u$ as
\begin{veqnarray}
  \nonumber \int_{-\infty}^u du' \ R^\nu_x(u',x,t)+\int d\xi\ \xi
  P^\nu =0,
\end{veqnarray}
can be used to derive from (\ref{eq:A.1}) an equation for $R^\nu$:
\begin{veqnarray}
  \label{eq:A.5}
  R^\nu_t&=&-u R^\nu_x-\int_{-\infty}^u du' \ R^\nu_x(u',x,t)\\
  &&+B_0 R^\nu_{uu}-\nu\bigl( \< u_{xx}|u\> R^\nu\bigr)_u.
\end{veqnarray}
For statistically homogeneous situation, $P^\nu_x=0$ in
(\ref{eq:A.1}).  Then, using $\int d\xi \, \xi P^\nu =0$ and
(\ref{eq:A.1}) leads to the following equations for $R^\nu$ and
$Q^\nu$:
\begin{veqnarray}
  \label{eq:A.6}
  R^\nu_t=B_0 R^\nu_{uu} -\nu \bigl(\< u_{xx}|u\> R^\nu\bigr)_{u}.
\end{veqnarray}
\begin{veqnarray}
  \label{eq:A.7}
  Q^\nu_t= \xi Q^\nu+\bigl(\xi^2 Q^\nu\bigr)_\xi+B_1Q^\nu_{\xi\xi}
  -\nu\bigl(\< \xi_{xx}|\xi\> Q^\nu\bigr)_{\xi}.
\end{veqnarray}

\ack We thank S. A. Boldyrev, R. H. Kraichnan, A. M. Polyakov, and Ya.
Sinai for very stimulating discussions.  The work of E is supported by
a Presidential Faculty Fellowship from the National Science
Foundation.  The work of Vanden Eijnden is supported by U.S.
Department of Energy Grant No. DE-FG02-86ER-53223.

\bibliographystyle{plain}

\begin{thebibliography}{9}
  
\bibitem{ave95b} Avellaneda, M., \textit{Statistical Properties of
    Shocks in Burgers Turbulence, II: Tail Probabilities for
    Velocities, Shock-Strengths and Rarefaction Intervals}, Commun.
  Math. Phys.  {\bf 169}, 45--59 (1995).
  
\bibitem{ave95a} Avellaneda, M., and E, W., \textit{Statistical
    Properties of Shocks in Burgers Turbulence}, Commun. Math. Phys.
  {\bf 172}, 13--38 (1995).
  
\bibitem{avrye94} Avellaneda, M., Ryan, R., and E, W., \textit{PDFs
    for velocity and velocity gradients in Burgers' turbulence}, Phys.
  Fluids {\bf 7}, 3067--3071 (1994).
  
\bibitem{bafako97} Balkovsky, E., Falkovich, G., Kolokolov, I., and
  Lebedev, V., \textit{Intermittency of Burgers' Turbulence}, Phys.
  Rev. Lett. {\bf 78}, 1452--1455 (1997).
  
\bibitem{blfege94} Blatter, G., Feigelman, M. V., Geshkenbein, V. B.,
  Larkin, A. I., and Vinokur, V. M., \textit{Vortices in
    high-temperature superconductors}, Rev. Modern Phys. {\bf 66},
  1125--1388 (1994).
  
\bibitem{bol97} Boldyrev, S. A., \textit{Velocity-difference
    probability density functions for Burgers turbulence}, Phys. Rev.
  E {\bf 55}, 6907--6910 (1997).
  
\bibitem{bol98} Boldyrev, S. A., \textit{Burgers turbulence,
    intermittency, and nonuniversality}, Phys. Plasmas {\bf 5},
  1681--1687 (1998).
  
\bibitem{bomepa95} Bouchaud, J.-P., M\'ezard, M., and Parisi, G.,
  \textit{Scaling and intermittency in Burgers turbulence}, Phys.
  Rev. E {\bf 52}, 3656--3674 (1995).
  
\bibitem{bome96} Bouchaud, J.-P., and M\'ezard, M., \textit{Velocity
    fluctuations in forced Burgers turbulence}, Phys. Rev. E {\bf 54},
  5116--5121 (1996).
  
\bibitem{chya95a} Chekhlov, A., and Yakhot, V., \textit{Kolmogorov
    turbulence in a random-force-driven Burgers equation}, Phys. Rev.
  E {\bf 51}, R2739--2742 (1995).
  
\bibitem{chya95b} Chekhlov, A., and Yakhot, V., \textit{Kolmogorov
    turbulence in a random-force-driven Burgers equation: Anomalous
    scaling and probability density functions}, Phys. Rev. E {\bf 51},
  R5681--5684 (1995).
  
\bibitem{chya96} Chekhlov, A., and Yakhot, V., \textit{Algebraic Tails
    of Probability Density Functions in the Random-Force-Driven
    Burgers Turbulence}, Phys. Rev. Lett. {\bf 77}, 3118--3121 (1996).
  
\bibitem{ekhma97} E, W., Khanin, K., Mazel, A., and Sinai, Ya. G.,
  \textit{Probability distributions functions for the random forced
    Burgers equation}, Phys. Rev. Lett. {\bf 78}, 1904--1907 (1997).
  
\bibitem{ekhma99} E, W., Khanin, K., Mazel, A., and Sinai, Ya. G.,
  \textit{Invariant measures for the random-forced Burgers equation},
  submitted to Ann.  Math.
  
\bibitem{eva99a} E, W., and Vanden Eijnden, E., \textit{Asymptotic
    theory for the probability density functions in Burgers
    turbulence}, chao-dyn/9901006, submitted to Phys. Rev. Lett.

  
\bibitem{eva99b} E, W., and Vanden Eijnden, E., \textit{On the
    statistical solution of Riemann equation and its implications for
    Burgers turbulence}, submitted to Phys. Fluids.
  
\bibitem{eva99c} E, W., and Vanden Eijnden, E., \textit{Another Note
    on forced Burgers turbulence}, chao-dyn/9901029.
  
\bibitem{fakole96} Falkovitch, G., Kolokolov, I., Lebedev, V., and
  Migdal, A., \textit{Instantons and intermittency}, Phys. Rev. E {\bf
    54}, 4896--4907 (1996).
  
\bibitem{feig83} Feigelman, M. V., \textit{One-dimensional periodic
    structures in a weak random potential}, Sov. Phys. JETP {\bf 52},
  555--561 (1980).

\bibitem{fofr83} Fournier, J. D, and Frisch, U., \textit{L'{\'e}quation de
  Burgers d{\'e}terministe et statistique}, J. M{\'e}c. Th{\'e}or.
  Appl.  {\bf 2}, 699--750 (1983).  

  
\bibitem{goxi92} Goodman, J., and Xin, Z., \textit{Viscous Limits for
    Piecewise Smooth Solutions to Systems of Conservation Laws}, Arch.
  Rational Mech. Anal. {\bf 121} 235--265 (1992).
  
\bibitem{got94} Gotoh, T., \textit{Inertial range statistics of
    Burgers turbulence}, Phys. Fluids {\bf 6}, 3985--3998 (1994).
  
\bibitem{gokr93} Gotoh, T., and Kraichnan, R. H., \textit{Statistics
    of decaying turbulence}, Phys. Fluids A {\bf 5}, 445--457 (1993).
  
\bibitem{gokr98} Gotoh, T. and Kraichnan, R. H., \textit{Steady-state
    Burgers turbulence with large-scale forcing}, Phys. Fluids {\bf
    10}, 2859--2866 (1998).
  
\bibitem{gumi96} Gurarie, V., and Migdal, A., \textit{Instantons in
    the Burgers equation}, Phys. Rev. E {\bf 54}, 4908--4914 (1996).
  
\bibitem{gusiau97} Gurbatov, S. N., Simdyankin, S. I., Aurell, E.
  Frisch, U., T{\'o}th, G., \textit{On the decay of Burgers
    turbulence}, {\bf 344}, 339--374 (1997).
  
\bibitem{kapazh86} Kardar, M., Parisi, G., and Zhang, Y. C.,
  \textit{Dynamic scaling of growing interfaces}, Phys. Rev. Lett.
  {\bf 56}, 889--892 (1986).
  
\bibitem{kra99} Kraichnan, R. H., \textit{Note on forced Burgers
    turbulence}, chao-dyn/9901023.
  
\bibitem{krsp92} Krug, J., and Spohn, H. \textit{Kinetic roughening of
    growing surfaces}, pp: 477--582 in: \textit{Solids far from
    equilibrium} (Godr{\`e}che, G. C., ed.), Cambridge University
  Press, England (1992).
  
\bibitem{lax57} Lax, P. D., \textit{Hyperbolic systems of conservation
    laws}, Comm. Pure Appl. Math. {\bf 10}, 537--566 (1957).
  
\bibitem{lax73} Lax, P. D., \textit{Hyperbolic Systems of Conservation
    Laws and the Mathematical Theory of Shock Waves}, SIAM,
  Philadelphia (1973).
  
\bibitem{liu99} Liu, T. P., \textit{}, CBMS regional conference
  lectures on shock wave theory (Atlanta, Augustus,1997), to appear.
  
\bibitem{pol95} Polyakov, A. M., \textit{Turbulence without pressure},
  Phys. Rev. E {\bf 52}, 6183--6188 (1995).

\bibitem{pol99} Polyakov, A. M., private communication.
  
\bibitem{ryav99} Ryan, R., and Avellaneda, M., \textit{The one-point
    statistics of viscous Burgers turbulence initialized with Gaussian
    data}, submitted to: Commun. Math. Phys.
  
\bibitem{shze89} Shandarin, S. F., Zeldovich, Ya. B., \textit{The
    large-scale structure of the universe: Turbulence, intermittency,
    structures in a self-graviting medium}, Rev. Mod. Phys. {\bf 61},
  185--220 (1989).
  
\bibitem{sin92} Sinai, Ya., \textit{Two results concerning asymptotic
    behavior of solutions of the Burgers equation with force}, J.
  Stat. Phys. {\bf 64}, 1--12 (1992).
  
\bibitem{sin96} Sinai, Ya., \textit{Burgers system driven by a
    periodic stochastic flow}, pp. 347--355 in: \textit{It\^{o}'s
    Stochastic Calculus and Probability Theory}, Springer-Verlag, New
  York, Berlin (1996).
  
\bibitem{smo83} Smoller, J., \textit{Shock Waves and Reaction
    Diffusion Equations}, Springer-Verlag, New York, Berlin (1983).
  
\bibitem{vedufr94} Vergassola, M., Dubrulle, B., Frisch, U., and
  Noullez, A., \textit{Burgers, equation, Devil's staircases and the
    mass distribution for large-scale structures}, Astron. Astrophys.
  {\bf 280}, 325--356 (1994).
  
\bibitem{vol67} Vol'pert, A., \textit{The space BV and quasilinear
    equations}, Math. USSR, Sb. {\bf 2} 225--267 (1967).
  

\end{thebibliography}

\end{document}